\documentclass{amsart}

\usepackage{amssymb, amsmath, amsthm}
\usepackage[fleqn,tbtags]{mathtools}
\usepackage{mathrsfs}
\usepackage{mathabx}
\usepackage{xcolor}
\usepackage{graphicx}

\usepackage[shortlabels]{enumitem}

\usepackage[active]{srcltx}
\usepackage{verbatim}
\usepackage{physics}
\usepackage{tikz}
\usepackage{tikz-cd}
\usepackage{quantikz}
\usetikzlibrary{arrows, positioning}

\usepackage{float}
\usepackage[colorlinks,linkcolor={blue},citecolor={blue},urlcolor={blue}]{hyperref}

\let\mathcal \undefined
\def\mathcal{\mathscr}

\let\leq\leqslant

\let\geq\geqslant

\theoremstyle{plain}
\newtheorem{theorem}{Theorem}[section]
\theoremstyle{remark}
\newtheorem{remark}[theorem]{Remark}
\newtheorem{example}[theorem]{Example}
\theoremstyle{plain}
\newtheorem{corollary}[theorem]{Corollary}
\newtheorem{lemma}[theorem]{Lemma}
\newtheorem{proposition}[theorem]{Proposition}
\newtheorem{definition}[theorem]{Definition}

\numberwithin{equation}{section}

\def\C{{\mathbb C}}

\renewcommand{\P}{{\mathbb P}}

\newcommand{\iprod}[2]{( #1|#2 )}

\newcommand{\calL}{\mathscr{L}}

\newcommand{\n}{\Vert}

\allowdisplaybreaks

\begin{document}

\title[Holevo classicalisation and context integration]{Holevo classicalisation and context integration: Global versus protocolwise integration}
\author{Marijn Waaijer \& Jan van Neerven}

\address{Delft Institute of Applied Mathematics\\
Delft University of Technology\\P.O. Box 5031\\2600 GA Delft\\The Netherlands}

\email{waaijermarijn@gmail.com, j.m.a.m.vanneerven@tudelft.nl}

\date{\today}


\begin{abstract} 
Suppose that a family of commutative measurement contexts is given, together with the corresponding classical probability laws. We study the problem under what conditions these contextwise descriptions can be combined into a single classical representation.
We distinguish two ways in which this can be done. In {\em global integration}, the prescribed observable algebras are required to fit into a single commutative context and the corresponding probability laws to arise as marginals of one joint law. In {\em protocolwise integration}, by contrast, the context label is retained as part of the description, and the contextwise laws appear as conditional distributions within a single model of the measurement protocol. We show that global integration involves two distinct notions of compatibility: operator-algebraic compatibility of the underlying observables and probabilistic compatibility of their classicalised laws. The former implies the latter, but the converse fails; under the topological hypotheses stated below, probabilistic compatibility is characterised by a Kellerer-type marginal criterion. Protocolwise integration, on the other hand, always exists and is represented operator-algebraically by a direct-sum construction. When global integration exists, the two constructions arise as reductions of a common commutative refinement. We illustrate the distinction in the Bell--CHSH experiment, where Fine's theorem relates global probabilistic integration to Bell locality and the CHSH inequalities. This shows that the obstruction revealed by Bell--CHSH concerns not classical representation as such, but classical representation preserving specified cross-context identifications.
\end{abstract}

\maketitle


\section{Introduction}\label{sec:introduction}

In \cite{van2026indiscernibility}, we studied the classical representation of a quantum system relative to a prescribed family of observables. A commutative algebra of observables determines an equivalence relation on quantum states according to the statistics accessible within that algebra, and the resulting Holevo space provides a classical representation of the corresponding measurement context. In this sense, every fixed commutative context admits its own classical description.

The final examples of \cite{van2026indiscernibility} point to a question that is not answered by this contextwise construction. In the setting of the Einstein-Podolsky-Rosen (EPR) experiment, the joint observable recording the pair of outcomes determines a joint probability distribution, whereas the separate perspectives of Alice and Bob determine only its marginals. Passing from the former to the latter forgets the correlation information contained in the joint distribution. In the setting of the experiment of Bell and Clauser, Horne, Shimony, and Holt (CHSH), the problem becomes more pronounced: different choices of measurement settings determine different commutative observable algebras, and these algebras are in general incompatible as subalgebras of the full quantum algebra. Thus, although each fixed context admits a classical representation, it does not follow that the resulting family of classical descriptions can itself be represented by one larger classical object.

This leaves open a natural integration problem. Suppose that a family of commutative measurement contexts is given, together with the corresponding classical probability laws. Under what conditions can these contextwise descriptions be combined into a single classical representation?
The formulation of this question conceals an important ambiguity, since there are two distinct ways in which contextwise descriptions may be combined:
\begin{enumerate}
\item As a \emph{global integration} of the prescribed contexts: the different marginal laws are required to arise as marginals of one joint probability law, while, at the operator-algebraic level, the different commutative observable algebras are required to fit into a single commutative context. In such a construction, the contextual separation of the original descriptions is overcome by identifying their common observables inside one larger classical object.
\item As a \emph{protocolwise integration} of the prescribed contexts. Instead of asking whether the contextwise laws are marginals of one context-independent joint law, one may form a classical model of the complete measurement protocol in which the choice of context is itself represented as an additional variable. The different contextwise distributions then occur as conditional distributions within a single probability space. At the operator-algebraic level, the corresponding construction likewise keeps the different contexts distinguished, so that observables occurring in different contexts need not first be identified inside one common commutative algebra. 
\end{enumerate}

The aim of the present paper is to separate these two integration procedures,
to compare them at both the operator-algebraic and probabilistic levels, and
to investigate their consequences. An analysis of the Bell-CHSH experiment shows that the existence of a global integration is non-trivial and gives rise to genuine compatibility problems that, in general, cannot be taken for granted. In contrast, protocolwise integration will be shown to exist without an additional compatibility condition.

The distinction between these two forms of
integration naturally suggests the more fundamental question, to be addressed prior to posing the standard
global-extension problem, namely: what justifies requiring global integration rather
than protocolwise integration? Standard mathematical formulations of
contextuality typically begin from an already specified measurement scenario,
consisting of a collection of measurements together with a family of compatible
contexts, as for example in the sheaf-theoretic approach of \cite{abramsky2011sheaf}. Such a specification
already determines how measurement occurrences belonging to different contexts
are organised into one common measurement structure. The subsequent question
is then whether the corresponding contextwise probability laws admit a global,
context-independent representation. In this sense, the structural choices that
make the global-extension problem possible are usually treated as prerequisites
of the formalism rather than as an organising question in their own right.

A notable exception is the approach of Dzhafarov and Kujala
\cite{dzhafarov2014contextuality,DzhafarovKujala2014}, who likewise take the
way in which contextual descriptions are combined to be an organising
question. Their analysis makes explicit that relations of cross-context
identity constitute additional requirements imposed on a probabilistic
coupling, rather than relations supplied by the individual contextwise
distributions themselves. The present paper brings this perspective to the
quantum--classical interface. We show that the distinction between global and
protocolwise integration can be formulated both at the level of operator
algebras and at the level of classical probability spaces, and investigate
how these two levels are related by classicalisation. Our aim is to show how
the two forms of integration give rise to two distinct commuting diagrams
connecting quantum observable structures with their classical probabilistic
representations, thereby relating the probabilistic perspective developed by
Dzhafarov and Kujala to the operator-algebraic structure of quantum
classicalisation.

\medskip

The paper is organised as follows. After briefly recalling the relevant results from \cite{van2026indiscernibility} and formalising the integration problem in Section~\ref{sec:previous-results}, we study global integration for a general family of commutative measurement contexts in Section~\ref{sec:global-context_integration}. We show that its existence is subject to distinct compatibility requirements at the probabilistic and operator-algebraic levels. Next, in Section~\ref{sec:protocol_wise_integration}, we show that protocolwise integration is always possible, at the cost of retaining the contextual structure in a larger classical description. In both cases, the passage from commutative observable algebras to classical probability models is organised by a commuting diagram, making precise how the two integration procedures interact with classicalisation.

We then specialise these general constructions to the Bell--CHSH experiment in Section~\ref{sec:application_to_bell_chsh}. For each fixed pair of measurement settings, the corresponding joint observable defines an ordinary commutative context and hence an ordinary classical probability law. The four resulting contextwise laws can always be integrated at the level of the experimental protocol by retaining the chosen settings as random variables. By contrast, their integration into a single probability law for all four potential outcomes is precisely the marginal compatibility problem characterised, in the Bell--CHSH case, by the Kellerer conditions. The Bell--CHSH example therefore makes concrete the distinction developed in the general theory: a family of classical contextwise descriptions always admits a unified classical description of the protocol, while it may fail to admit a single classical description in which the contextual distinctions have been removed. The implications of this distinction and its relation to broader approaches to contextuality and classical representation are discussed in the final section.

\section{Holevo classicalisation and the integration problem}
\label{sec:previous-results}

In this section we recall the main results from \cite{van2026indiscernibility}. Throughout, \(H\) denotes a separable complex Hilbert space with inner product \(\iprod{\cdot}{\cdot}\), linear in its first variable and sesquilinear in its second, and norm \(\n\cdot\n\) given by
\[
    \n x\n^2=\iprod{x}{x},
    \qquad x\in H.
\]
We write \(\calL(H)\) for the bounded linear operators on \(H\), \(\mathcal{S}(H)\) for the normal states, identified with the positive trace-class operators of trace one, and \(\mathcal{P}(H)\) for the pure normal states. Every norm-one vector \(h\in H\) determines the rank-one pure state
\[
    \rho_h:=\ket{h}\!\bra{h},
\]
and every pure normal state has this form. Moreover, \(\rho_h=\rho_{h'}\) if and only if \(h=ch'\) for some \(c\in\C\) with \(|c|=1\). When no confusion can arise, we write \([h]_{\mathcal M}\) for the indiscernibility class of the pure state \(\rho_h\).
The states considered in this paper are always normal, and we will omit this adjective when no confusion is likely to arise. Thus, the term `(pure) state' is used as short-hand for a normal (pure) state.

An {\em algebra of observables} is a von Neumann algebra \(\mathcal{M}\subseteq\mathcal{L}(H)\). We shall refer to a commutative von Neumann algebra as a \emph{measurement context}.

\subsection{Holevo spaces and classicalisation}

Let \(\mathcal{M}\subseteq\mathcal{L}(H)\) be a von Neumann algebra. Two states
\(\rho,\rho'\in\mathcal{S}(H)\) are called \emph{indiscernible with respect to \(\mathcal{M}\)} if
\[
    \operatorname{tr}(a\rho)
    =
    \operatorname{tr}(a\rho')
    \qquad
    \text{for all } a\in\mathcal{M}.
\] 
We write
\[
    \rho\sim_{\mathcal{M}}\rho'
\]
for this equivalence relation. 
The corresponding quotient
\[
    \mathcal{S}(H)/\mathcal{M}
\]
is the {\em Holevo space} associated with \(\mathcal{M}\). Restricting the same equivalence relation to pure states gives the {\em pure Holevo space}
\[
    \mathcal{P}(H)/\mathcal{M}.
\]
The Holevo space may be identified with the normal state space of the observable algebra itself. Indeed, the mapping
\[
    [\rho]_{\mathcal{M}}
    \longmapsto
    \omega_\rho,
    \qquad
    \omega_\rho(a)
    :=
    \operatorname{tr}(a\rho),
\]
defines an affine bijection
\[
    \mathcal{S}(H)/\mathcal{M}
    \cong
    \mathcal{S}(\mathcal{M}),
\]
where \(\mathcal{S}(\mathcal{M})\) denotes the set of normal states on \(\mathcal{M}\); see
\cite[Proposition~4.2]{van2026indiscernibility}. Thus the Holevo space retains precisely the distinctions between states that are accessible through the prescribed algebra of observables.

The same construction may be formulated directly for a projection-valued measure (PVM). Let
\[
    P:\Sigma\longrightarrow\operatorname{Proj}(H)
\]
be a projection-valued measure on a measurable space \((X,\Sigma)\), which we think of as the {\em outcome space} of the observable encoded by \(P\), and let
\[
    \mathcal{M}_P
    :=
    W^*\bigl(P(E):E\in\Sigma\bigr)
\]
be the von Neumann algebra generated by its range. In this setting, two states are called {\em indiscernible with respect to \(P\)} if
\[
    \operatorname{tr}\bigl(P(E)\rho\bigr)
    =
    \operatorname{tr}\bigl(P(E)\rho'\bigr)
    \qquad
    \text{for all } E\in\Sigma.
\]
By \cite[Proposition~3.5]{van2026indiscernibility},
\[
    \rho\sim_P\rho'
    \quad\Longleftrightarrow\quad
    \rho\sim_{\mathcal{M}_P}\rho',
\]
which shows that the two indiscernibility notions coincide.
Consequently, the probability law of a PVM and the observable algebra generated by that PVM determine the same Holevo space.

The central representation result of \cite{van2026indiscernibility} may be recalled as follows.

\begin{theorem}[Holevo classicalisation]
\label{thm:holevo-classicalisation}
Let \(\mathcal{M}\subseteq \calL(H)\) be a von Neumann algebra. 
The pure Holevo space
\(X_{\mathcal{M}}=\mathcal{P}(H)/\mathcal{M}\) carries a natural Hausdorff topology such that, for every \(a\in\mathcal{M}\), the function
\[
    \widehat{a}:X_{\mathcal{M}}\longrightarrow\mathbb C,
    \qquad
    \widehat{a}([h]_{\mathcal{M}})
    =
    \iprod{ah}{h},
\]
is bounded and continuous. Moreover, the assignment
\[
    \mathcal{M}\longrightarrow C_{\rm b}(X_{\mathcal{M}}),
    \qquad
    a\longmapsto\widehat{a},
\]
is injective.
\end{theorem}

This is \cite[Theorem~6.1]{van2026indiscernibility}; we refer to that paper for the construction of the topology and the proof. An analogous formulation holds for mixed states, now taking
\[
    X_{\mathcal{M}}
    =
    \mathcal{S}(H)/\mathcal{M}
\]
and
\[
    \widehat{a}([\rho]_{\mathcal{M}})
    =
    \operatorname{tr}(a\rho).
\]
 In both versions, the theorem assigns to each quantum observable \(a\in \mathcal{M}\) a classical observable (i.e., a pointwise defined function) \(\widehat a\) defined on a classical state space (i.e., a topological Hausdorff space \(X_{\mathcal{M}}\)).

Theorem~\ref{thm:holevo-classicalisation} applies to arbitrary von Neumann algebras \(\mathcal{M}\), including noncommutative ones. In the present paper, however, the main role will be played by the stronger probabilistic interpretation available when \(\mathcal{M}\) is commutative. In that case, \cite[Theorem~5.6]{van2026indiscernibility} asserts that the pure and mixed Holevo spaces coincide up to affine isomorphism,
\[
    \mathcal{P}(H)/\mathcal{M}
    \cong
    \mathcal{S}(H)/\mathcal{M},
\]
and the classicalisation may be represented in terms of ordinary probability laws associated with a joint measurement.

More precisely, since \(H\) is separable, \(\mathcal M\) is generated by the PVM of a single self-adjoint element; see the proof of \cite[Theorem~5.6]{van2026indiscernibility}. Let \(P\) be such a generating PVM. For every state \(\rho\in\mathcal{S}(H)\), define
\[
    \mu^\rho_P(E)
    :=
    \operatorname{tr}\bigl(P(E)\rho\bigr),
    \qquad
    E\in\Sigma.
\]
Then \(\mu^\rho_P\) is a probability measure on the outcome space \((X,\Sigma)\), and
\[
    [\rho]_{\mathcal{M}}
    =
    [\rho']_{\mathcal{M}}
    \quad\Longleftrightarrow\quad
    \mu^\rho_P
    =
    \mu^{\rho'}_P.
\]
Thus, for a commutative measurement context, the corresponding Holevo state is completely represented by the ordinary probability law of its joint observable. In the case of a finite family of commuting normal observables, this representation may equivalently be formulated on their joint spectrum; see \cite[Corollary~5.7]{van2026indiscernibility}.

For the purposes of the present paper, the basic ``unit'' of classicalisation is therefore the passage
\[
    (\mathcal{M},\rho)
    \quad\longmapsto\quad
    \mu^\rho_{\mathcal{M}},
\]
from a commutative measurement context and a quantum state to the associated classical probability law.

\subsection{The integration problem}\label{sec:integration_problem}

We consider a measurement set-up which has been organised into a family of contexts, with specified identifications of measurements occurring in different contexts. This can be mathematically formulated as follows. For every \(i\in I\), where \(I\) is a finite index set, let
\[
    P_i:\Sigma_i\longrightarrow\operatorname{Proj}(H)
\]
be a projection-valued measure on a measurable outcome space \((X_i,\Sigma_i)\). Let
\[
    \mathfrak{C}\subseteq 2^I
\]
be a family of subsets of \(I\), called \emph{contexts}, such that
\[
    \bigcup_{C\in\mathfrak{C}} C = I,
\]
and such that for every context \(C\in\mathfrak{C}\) the PVMs
\[
    \{P_i:i\in C\}
\]
commute pairwise.

For every \(C\in\mathfrak{C}\), the commuting family \((P_i)_{i\in C}\) admits a joint PVM
\[
    P_C:
    \bigotimes_{i\in C}\Sigma_i
    \longrightarrow
    \operatorname{Proj}(H)
\]
on the product outcome space
\[
    X_C
    :=
    \prod_{i\in C}X_i.
\]
On measurable rectangles it is characterised by
\[
    P_C\Bigl(\prod_{i\in C}E_i\Bigr)
    =
    \prod_{i\in C}P_i(E_i).
\]
The corresponding observable algebra
\[
    \mathcal{M}_C
    :=
    W^*\bigl(P_i(E):i\in C,\ E\in\Sigma_i\bigr)
\]
is commutative and therefore defines a measurement context in the above sense.

The range of \(P_C\) generates \(\mathcal M_C\). Indeed, its rectangle projections belong to \(\mathcal M_C\), and a \(\pi\)-\(\lambda\) argument gives \(P_C(E)\in\mathcal M_C\) for every \(E\in\bigotimes_{i\in C}\Sigma_i\); conversely, each \(P_i(E_i)\), \(i\in C\), is a marginal projection of \(P_C\).

Fix a state \(\rho\in\mathcal{S}(H)\). Classicalisation of each context  \(C\in\mathfrak{C}\) gives a probability law
\[
    \mu^\rho_C(E)
    :=
    \operatorname{tr}\bigl(P_C(E)\rho\bigr),
    \qquad
    E\in\bigotimes_{i\in C}\Sigma_i.
\]
We therefore obtain two parallel families,
\[
    \{\mathcal{M}_C\}_{C\in\mathfrak{C}}
    \qquad\text{and}\qquad
    \{\mu^\rho_C\}_{C\in\mathfrak{C}},
\]
related contextwise by classicalisation.

These contextwise laws are compatible on overlaps. Indeed, the
global indexing already prescribes which measurement occurrences in different
contexts are to be identified: if \(i\in C\cap D\), the same PVM \(P_i\)
occurs in both contexts. Thus, if \(C,D\in\mathfrak{C}\) have non-empty intersection and
\[
    \pi^C_{C\cap D}:X_C\longrightarrow X_{C\cap D},
    \qquad
    \pi^D_{C\cap D}:X_D\longrightarrow X_{C\cap D}
\]
denote the canonical coordinate projections, we have the equality of image measures
\[
    \bigl(\pi^C_{C\cap D}\bigr)_{\#}\mu^\rho_C
    =
    \bigl(\pi^D_{C\cap D}\bigr)_{\#}\mu^\rho_D.
\]
Both sides are equal to the probability law associated with the common
subfamily
\[
    \{P_i:i\in C\cap D\}.
\]
Thus the classical descriptions supplied by the individual contexts agree
wherever the corresponding measurement descriptions overlap. This automatic
overlap compatibility is therefore relative to the cross-context
identifications already encoded by the global indexing; the significance of
this prior identification will become important when global and protocolwise
integration are compared below.

The local compatibility on overlaps, however, does not yet provide a single classical representation of the full family. The construction above gives only
\[
    \{\mathcal{M}_C\}_{C\in\mathfrak{C}}
    \xrightarrow{\ \mathrm{classicalisation}\ }
    \{\mu^\rho_C\}_{C\in\mathfrak{C}},
\]
that is, a family of classical representations indexed by the prescribed measurement contexts. 
The problem addressed in the present paper is how this family is to be integrated into a more comprehensive description of the experiment. In the language of the Holevo-space construction, the question can now be formulated as follows: under what requirements, and under what conditions, can the contextwise observable algebras and their corresponding probability laws be integrated into a larger description such that the diagram in Figure~\ref{fig:general-integration-problem} commutes?
As we shall see, the answer depends on how the contextual structure is treated.

\begin{figure}[ht]
\centering
\resizebox{0.98\linewidth}{!}{%
\begin{tikzpicture}[
    node distance=2.8cm and 3.2cm,
    box/.style={
        draw,
        align=center,
        minimum width=4.4cm,
        minimum height=1.2cm
    },
    arrow/.style={
        -{Latex[length=2.5mm]},
        thick
    }
]

\node[box] (contexts) {
    Prescribed measurement \\ contexts\\[1mm]
    $\{\mathcal{M}_C\}_{C\in\mathfrak{C}}$
};

\node[box, right=of contexts] (laws) {
    Prescribed contextwise  \\ laws\\[1mm]
    $\{\mu_C^\rho\}_{C\in\mathfrak{C}}$
};

\node[box, below=of contexts] (integrated-algebra) {
    Integrated observable \\ description\\[1mm]
    $\mathcal{M}_{\mathrm{int}}$
};

\node[box, right=of integrated-algebra] (integrated-law) {
    Integrated classical \\ description\\[1mm]
    $\mu_{\mathrm{int}}^\rho$
};

\draw[arrow]
    (contexts) --
    node[above, align=center] {contextwise\\classicalisation}
    (laws);

\draw[arrow]
    (contexts) --
    node[left, align=right] {integration of\\contexts}
    (integrated-algebra);

\draw[arrow]
    (laws) --
    node[right, align=left] {integration of\\contextwise laws}
    (integrated-law);

\draw[arrow]
    (integrated-algebra) --
    node[below] {classicalisation}
    (integrated-law);

\end{tikzpicture}%
}
\caption{The general integration problem. Starting from a family of commutative measurement contexts and their contextwise classicalisations, one asks for an integrated observable description and a corresponding integrated classical description such that the diagram commutes. The meaning of the vertical integration procedures depends on how the contextual structure is treated.}
\label{fig:general-integration-problem}
\end{figure}

\section{Global-context integration}
\label{sec:global-context_integration}

We begin with the first interpretation of the integration problem formulated in the preceding section: the global integration of the prescribed measurement contexts. At the operator-algebraic level, this requires the commutative observable algebras \(\{\mathcal{M}_C\}_{C\in\mathfrak{C}}\) to arise as subalgebras of one common commutative context. At the probabilistic level, it requires the associated contextwise laws \(\{\mu_C^\rho\}_{C\in\mathfrak{C}}\) to arise as marginals of one probability law on the full outcome space. We first show that, whenever the observable contexts admit such a common commutative envelope, Holevo classicalisation carries this operator-algebraic integration to a global probabilistic integration, and the corresponding instance of Figure~\ref{fig:general-integration-problem} commutes. We then formulate the operator-algebraic and probabilistic requirements as two distinct compatibility conditions and study the ways in which global integration may fail. In particular, although a common commutative context always induces a global coupling of the associated probability laws, the converse need not hold: passing from observable algebras to their classical probability laws preserves global integration but does not, in general, reflect it. We conclude the section by collecting these results in the global-integration diagram of Figure~\ref{fig:global_classicalisation_diagram}.

\subsection{Global integration in a common commutative context}
\label{sec:global_commutative_context}

We first identify three equivalent formulations of global integration at the operator-algebraic level.

\begin{proposition}[Common commutative contexts and joint observables]
\label{prop:common-context-joint-pvm}
Let
\[
    \mathcal{M}_{\mathrm{env}}
    :=
    W^*\Bigl(
        \bigcup_{C\in\mathfrak{C}}\mathcal{M}_C
    \Bigr)
    =
    W^*\bigl(
        P_i(E):
        i\in I,\ E\in\Sigma_i
    \bigr)
\]
be the enveloping observable algebra. The following assertions are equivalent:
\begin{enumerate}[\rm(1)]

    \item\label{it:common-context-joint-pvm1} The enveloping observable algebra
    \(\mathcal{M}_{\mathrm{env}}\) is commutative.
    
    \item\label{it:common-context-joint-pvm2} The family
    \(
        \{\mathcal{M}_C\}_{C\in\mathfrak{C}}
    \)
    is contained in a common commutative von Neumann algebra
    \(\mathcal{N}\subseteq\mathcal{L}(H)\).

    \item\label{it:common-context-joint-pvm3} The family of PVMs
    \(
        \{P_i\}_{i\in I}
    \)
    admits a joint PVM
    \[
        P_I:
        \bigotimes_{i\in I}\Sigma_i
        \longrightarrow
        \operatorname{Proj}(H)
    \]
    on
    \[
        X_I:=\prod_{i\in I}X_i,
    \]
    characterised on measurable rectangles by
    \[
        P_I\Bigl(\prod_{i\in I}E_i\Bigr)
        =
        \prod_{i\in I}P_i(E_i).
    \]
\end{enumerate}

Whenever these equivalent conditions hold, the joint PVM \(P_I\) is unique and
\[
    \mathcal{M}_{\mathrm{env}}
    =
    W^*\bigl(
        P_I(E):
        E\in\textstyle\bigotimes_{i\in I}\Sigma_i
    \bigr).
\]
Moreover, for every context \(C\in\mathfrak{C}\), the contextwise joint PVM
\[
    P_C:
    \bigotimes_{i\in C}\Sigma_i
    \longrightarrow
    \operatorname{Proj}(H)
\]
is the marginal of \(P_I\):
\[
    P_C(E)
    =
    P_I\bigl(\pi_C^{-1}(E)\bigr),
    \qquad
    E\in\bigotimes_{i\in C}\Sigma_i,
\]
where
\[
    \pi_C:X_I\longrightarrow X_C
\]
is the canonical coordinate projection.
\end{proposition}

\begin{proof}
Suppose first that \(\mathcal{M}_{\mathrm{env}}\) is commutative. Then it is itself a common commutative von Neumann algebra containing all the prescribed contexts. Thus
\ref{it:common-context-joint-pvm1} implies \ref{it:common-context-joint-pvm2}.

Conversely, suppose that the prescribed contexts are contained in a common commutative von Neumann algebra \(\mathcal{N}\). Since \(\mathcal{N}\) contains every
\(\mathcal{M}_C\), it contains the von Neumann algebra generated by their union:
\[
    \mathcal{M}_{\mathrm{env}}
    \subseteq
    \mathcal{N}.
\]
It follows that \(\mathcal{M}_{\mathrm{env}}\) is commutative. Thus
\ref{it:common-context-joint-pvm2} implies \ref{it:common-context-joint-pvm1}.

Assume now that \(\mathcal{M}_{\mathrm{env}}\) is commutative. Since the range of every \(P_i\) is contained in \(\mathcal{M}_{\mathrm{env}}\), all projections
\[
    P_i(E),
    \qquad
    i\in I,\ E\in\Sigma_i,
\]
commute. The joint spectral theorem for commuting PVMs therefore gives a unique joint PVM
\[
    P_I:
    \bigotimes_{i\in I}\Sigma_i
    \longrightarrow
    \operatorname{Proj}(H)
\]
satisfying
\[
    P_I\Bigl(\prod_{i\in I}E_i\Bigr)
    =
    \prod_{i\in I}P_i(E_i)
\]
on measurable rectangles. Thus
\ref{it:common-context-joint-pvm1} implies \ref{it:common-context-joint-pvm3}.

Finally, suppose that such a joint PVM \(P_I\) exists. For every \(i\in I\) and
\(E\in\Sigma_i\),
\[
    P_i(E)
    =
    P_I\bigl(\pi_i^{-1}(E)\bigr),
\]
where \(\pi_i:X_I\to X_i\) is the \(i\)-th coordinate projection. Hence every
\(P_i(E)\) belongs to the commutative von Neumann algebra generated by the range of \(P_I\). It follows that
\[
    \mathcal{M}_{\mathrm{env}}
    \subseteq
    W^*\bigl(
        P_I(E):
        E\in\textstyle\bigotimes_{i\in I}\Sigma_i
    \bigr),
\]
and therefore \(\mathcal{M}_{\mathrm{env}}\) is commutative. Thus
\ref{it:common-context-joint-pvm3} implies \ref{it:common-context-joint-pvm1}.

For the reverse inclusion, put \(\mathcal N:=\mathcal M_{\mathrm{env}}\) and let \(\mathscr D\) be the collection of all sets \(E\in\bigotimes_{i\in I}\Sigma_i\) for which \(P_I(E)\in\mathcal N\). The class \(\mathscr D\) is a Dynkin system: it contains \(X_I\), is closed under complements, and is closed under countable disjoint unions because \(P_I\) is countably additive in the strong operator topology and \(\mathcal N\) is strongly closed. Every measurable rectangle belongs to \(\mathscr D\), since its projection is a finite product of projections \(P_i(E_i)\) belonging to \(\mathcal N\). The rectangles form a \(\pi\)-system generating \(\bigotimes_{i\in I}\Sigma_i\); hence the \(\pi\)-\(\lambda\) theorem gives \(\mathscr D=\bigotimes_{i\in I}\Sigma_i\). Therefore the full range of \(P_I\) is contained in \(\mathcal M_{\mathrm{env}}\), and consequently
\[
    \mathcal{M}_{\mathrm{env}}
    =
    W^*\bigl(
        P_I(E):
        E\in\textstyle\bigotimes_{i\in I}\Sigma_i
    \bigr).
\]

It remains to identify the contextwise PVMs as marginals of \(P_I\). Fix
\(C\in\mathfrak{C}\) and define
\[
    Q_C(E)
    :=
    P_I\bigl(\pi_C^{-1}(E)\bigr),
    \qquad
    E\in\bigotimes_{i\in C}\Sigma_i.
\]
Then \(Q_C\) is a PVM on \(X_C\). On a measurable rectangle
\(E=\prod_{i\in C}E_i\)
one has
\[
\begin{aligned}
    Q_C(E)
    =
    P_I\Bigl(
        \prod_{i\in C}E_i
        \times
        \prod_{i\notin C}X_i
    \Bigr)
    =
    \prod_{i\in C}P_i(E_i)
    \prod_{i\notin C}P_i(X_i)
    =
    \prod_{i\in C}P_i(E_i)
    =
    P_C(E).
\end{aligned}
\]
By uniqueness of the joint PVM associated with the commuting family
\((P_i)_{i\in C}\), it follows that \(Q_C=P_C\). Therefore
\[
    P_C(E)
    =
    P_I\bigl(\pi_C^{-1}(E)\bigr)
\]
for every \(E\in\bigotimes_{i\in C}\Sigma_i\).
\end{proof}

We may now state the main result of this subsection.

\begin{theorem}[Global classicalisation in the commutative case]
\label{thm:global_classicalisation_commutative}
Suppose that the enveloping observable algebra
\[
    \mathcal{M}_{\mathrm{env}}
    =
    W^*\Bigl(
        \bigcup_{C\in\mathfrak{C}}\mathcal{M}_C
    \Bigr)
\]
is commutative. Then, for every state
\(\rho\in\mathcal{S}(H)\), the probability measure
\[
    \mu_I^\rho(E)
    :=
    \operatorname{tr}\bigl(P_I(E)\rho\bigr),
    \qquad
    E\in\bigotimes_{i\in I}\Sigma_i,
\]
has the prescribed contextwise laws as its marginals:
\[
    (\pi_C)_{\#}\mu_I^\rho
    =
    \mu_C^\rho
    \qquad
    \text{for every }C\in\mathfrak{C}.
\]
Consequently, a common commutative context for the prescribed observable algebras induces a global probabilistic integration of their contextwise classicalisations.
\end{theorem}

\begin{proof}
Fix \(C\in\mathfrak{C}\) and
\(E\in\bigotimes_{i\in C}\Sigma_i\). By the definition of the push-forward measure and the marginal identity for the joint PVM,
\[
\begin{aligned}
    \bigl((\pi_C)_{\#}\mu_I^\rho\bigr)(E)
    =
    \mu_I^\rho\bigl(\pi_C^{-1}(E)\bigr)
    =
    \operatorname{tr}\bigl(
        P_I(\pi_C^{-1}(E))\rho
    \bigr)
    =
    \operatorname{tr}\bigl(P_C(E)\rho\bigr)
    =
    \mu_C^\rho(E).
\end{aligned}
\]
Since this holds for every \(E\in\bigotimes_{i\in C}\Sigma_i\), we obtain
\[
    (\pi_C)_{\#}\mu_I^\rho=\mu_C^\rho.
\]
As \(C\in\mathfrak{C}\) was arbitrary, \(\mu_I^\rho\) is a global coupling of the full family
\(\{\mu_C^\rho\}_{C\in\mathfrak{C}}\).
\end{proof}

The theorem can also be formulated directly at the level of Holevo spaces. Since
\[
    \mathcal{M}_C\subseteq\mathcal{M}_{\mathrm{env}},
\]
indiscernibility with respect to the enveloping algebra implies indiscernibility with respect to every prescribed context:
\[
    \rho\sim_{\mathcal{M}_{\mathrm{env}}}\rho'
    \quad\Longrightarrow\quad
    \rho\sim_{\mathcal{M}_C}\rho'.
\]
Hence there is a well-defined affine restriction map
\[
    r_C:
    \mathcal{S}(H)/\mathcal{M}_{\mathrm{env}}
    \longrightarrow
    \mathcal{S}(H)/\mathcal{M}_C,
    \qquad
    r_C\bigl([\rho]_{\mathcal{M}_{\mathrm{env}}}\bigr)
    :=
    [\rho]_{\mathcal{M}_C}.
\]
Under the identification of Holevo spaces with normal state spaces, this is simply the restriction
\[
    \omega
    \longmapsto
    \omega|_{\mathcal{M}_C}.
\]

Let
\[
    \Phi_I:
    \mathcal{S}(H)/\mathcal{M}_{\mathrm{env}}
    \longrightarrow
    \mathsf{P}(X_I),
    \qquad
    \Phi_I\bigl([\rho]_{\mathcal{M}_{\mathrm{env}}}\bigr)
    :=
    \mu_I^\rho,
\]
and
\[
    \Phi_C:
    \mathcal{S}(H)/\mathcal{M}_C
    \longrightarrow
    \mathsf{P}(X_C),
    \qquad
    \Phi_C\bigl([\rho]_{\mathcal{M}_C}\bigr)
    :=
    \mu_C^\rho,
\]
where \(\mathsf{P}(X)\) denotes the set of probability measures on the relevant measurable space. Theorem~\ref{thm:global_classicalisation_commutative} states that, for every \(C\in\mathfrak{C}\), the diagram
\[
\begin{array}{ccc}
\mathcal{S}(H)/\mathcal{M}_{\mathrm{env}}
&
\xrightarrow{\quad\Phi_I\quad}
&
\mathsf{P}(X_I)
\\[3mm]
{\scriptstyle r_C}\downarrow
&&
\downarrow{\scriptstyle(\pi_C)_{\#}}
\\[3mm]
\mathcal{S}(H)/\mathcal{M}_C
&
\xrightarrow{\quad\Phi_C\quad}
&
\mathsf{P}(X_C)
\end{array}
\]
commutes. Equivalently,
\[
    \Phi_C\circ r_C
    =
    (\pi_C)_{\#}\circ\Phi_I.
\]

Thus restriction from the global commutative observable algebra to a prescribed context is carried by Holevo classicalisation to marginalisation of the corresponding probability law. In this sense, classicalisation preserves global integration.

\subsection{The two compatibility conditions}
\label{sec:global_compatibility_conditions}

The preceding subsection shows that a common commutative context induces a global coupling of the associated contextwise probability laws. We now formulate the two compatibility conditions involved in this implication separately. The first is a structural condition on the prescribed observables, whereas the second is a condition on the probability laws generated by a fixed state.

\subsubsection{Operator-algebraic compatibility}
\label{sec:operator_algebraic_compatibility}

\begin{definition}[Operator-algebraic compatibility]
\label{def:operator_algebraic_compatibility}
The family of prescribed measurement contexts
\(
    \{\mathcal{M}_C\}_{C\in\mathfrak{C}}
\)
is called \emph{operator-algebraically compatible} if it is contained in a common commutative von Neumann algebra.
\end{definition}

By Proposition~\ref{prop:common-context-joint-pvm}, operator-algebraic compatibility is equivalent to commutativity of the enveloping observable algebra
\[
    \mathcal{M}_{\mathrm{env}}
    =
    W^*\Bigl(
        \bigcup_{C\in\mathfrak{C}}\mathcal{M}_C
    \Bigr).
\]

\subsubsection{Probabilistic compatibility and the Kellerer criterion}
\label{sec:probabilistic_compatibility}

We next formulate the corresponding condition on the contextwise probability laws. For the remainder of this subsection, assume that each \(X_i\) is a locally compact second-countable Hausdorff space and that
\(
    \Sigma_i=\mathcal{B}(X_i)
\) is its Borel \(\sigma\)-algebra. Then every finite product \(X_J=\prod_{i\in J}X_i\) is again locally compact and second countable, and its Borel \(\sigma\)-algebra agrees with \(\bigotimes_{i\in J}\Sigma_i\). Assume further that the contextwise laws
\(
    \mu_C\in\mathsf{P}_{\mathrm{R}}(X_C)
\),
\(C\in\mathfrak{C}\),
are Radon probability measures. Here
\[
    X_C:=\prod_{i\in C}X_i,
    \qquad
    X_I:=\prod_{i\in I}X_i,
\]
and \(\mathsf{P}_{\mathrm{R}}(X)\) denotes the set of Radon probability measures on \(X\). 

\begin{definition}[Probabilistic compatibility]
\label{def:probabilistic_compatibility}
A family of contextwise laws
\(
    \{\mu_C\}_{C\in\mathfrak{C}}
\)
is called \emph{probabilistically compatible} if there exists a Radon probability measure
\(
    \mu_I\in\mathsf{P}_{\mathrm{R}}(X_I)
\)
such that
\[
    (\pi_C)_{\#}\mu_I=\mu_C
    \qquad
    \text{for every }C\in\mathfrak{C},
\]
where
\(
    \pi_C:X_I\longrightarrow X_C
\)
is the canonical coordinate projection. Such a measure \(\mu_I\) will be called a
\emph{global coupling} of the family.
\end{definition}

\begin{remark}
In the finite measurement scenarios considered in the contextuality
literature, this is precisely the global-section condition for an empirical
model in the sense of \cite{abramsky2011sheaf}.
\end{remark}

The overlap consistency established in Section~\ref{sec:previous-results} is necessary for probabilistic compatibility. It is not, in general, sufficient: even when the prescribed laws agree on all common subcontexts, they need not arise as marginals of a single probability law. The complete obstruction is described by the following dual criterion.

For a locally compact Hausdorff space \(X\), let
\(
    C_0(X;\mathbb{R})
\)
denote the real-valued continuous functions vanishing at infinity, and let \(\mathsf M(X)\) denote the finite signed Radon measures on \(X\), identified with \(C_0(X;\mathbb R)^*\) by the Riesz representation theorem. We also let
\(
    C_{\rm b}(X;\mathbb{R})
\)
denote the bounded real-valued continuous functions.

\begin{lemma}[Relative closedness of the marginal image]
\label{lem:marginal_image_closed}
Let \(I\) and \(\mathfrak{C}\) be finite, with
\(
    \bigcup_{C\in\mathfrak{C}}C=I,
\)
and let each \(X_i\) be locally compact second-countable Hausdorff. Define
\[
    \mathsf{E}
    :=
    \prod_{C\in\mathfrak{C}}\mathsf{M}(X_C),
\]
where each space of signed Radon measures is equipped with its weak-* topology, and let
\[
    T:
    \mathsf{P}_{\mathrm{R}}(X_I)
    \longrightarrow
    \mathsf{E},
    \qquad
    T\nu
    :=
    \bigl((\pi_C)_{\#}\nu\bigr)_{C\in\mathfrak{C}}.
\]
This mapping is well defined: each push-forward is a Borel probability measure on the locally compact second-countable space \(X_C\), and hence is Radon.
Put
\[
    \mathsf{K}
    :=
    T\bigl(\mathsf{P}_{\mathrm{R}}(X_I)\bigr).
\]
Then
\[
    \overline{\mathsf{K}}^{\,\mathsf{E}}
    \cap
    \prod_{C\in\mathfrak{C}}
    \mathsf{P}_{\mathrm{R}}(X_C)
    =
    \mathsf{K}.
\]
Equivalently, if a net
\[
    T\nu_\alpha
    \longrightarrow
    (\mu_C)_{C\in\mathfrak{C}}
\]
in \(\mathsf{E}\), where every \(\mu_C\) is a Radon probability measure, then the family
\(
    (\mu_C)_{C\in\mathfrak{C}}
\)
admits a global coupling.
\end{lemma}

\begin{proof}
For each \(i\in I\), let \(X_i^+\) denote the one-point compactification of \(X_i\); when \(X_i\) is already compact, we take \(X_i^+=X_i\). Put
\[
    K_I:=\prod_{i\in I}X_i^+,
    \qquad
    K_C:=\prod_{i\in C}X_i^+.
\]
Since \(I\) is finite, \(K_I\) is compact Hausdorff, and \(X_I\) is an open subspace of \(K_I\). Here and below, \(\mathsf P(K)\) denotes the Radon probability measures on a compact Hausdorff space \(K\).

Regard each \(\nu_\alpha\) as a Radon probability measure
\(
    \overline{\nu}_\alpha
    \in
    \mathsf{P}(K_I)
\)
by extending it by zero outside \(X_I\). This extension is Radon because \(X_I\) is open in \(K_I\) and \(\nu_\alpha\) is a finite Radon measure. Since
\(\mathsf{P}(K_I)\) is weak-* compact, there exists a subnet, which we continue to denote by
\((\overline{\nu}_\alpha)\), and a probability measure
\(
    \overline{\nu}\in\mathsf{P}(K_I)
\)
such that
\[
    \overline{\nu}_\alpha
    \longrightarrow
    \overline{\nu}
\]
weak-* on \(K_I\).

For each \(C\in\mathfrak{C}\), the coordinate projection
\(
    \pi_C:X_I\longrightarrow X_C
\)
extends to a continuous projection
\(
    \overline{\pi}_C:K_I\longrightarrow K_C.
\)
Let
\[
    \overline{\nu}_C
    :=
    (\overline{\pi}_C)_{\#}\overline{\nu}.
\]

Fix \(C\in\mathfrak{C}\) and
\(f\in C_0(X_C;\mathbb{R})\). 
Since \(f\in C_0(X_C;\mathbb{R})\), its extension by zero,
\[
    \widetilde f(x)
    :=
    \begin{cases}
        f(x), & x\in X_C,\\
        0, & x\in K_C\setminus X_C,
    \end{cases}
\]
belongs to \(C(K_C;\mathbb{R})\). Indeed, for every
\(\varepsilon>0\), the set
\(
    \{x\in X_C:|f(x)|\geq\varepsilon\}
\)
is compact in \(X_C\), hence closed in \(K_C\); this proves continuity
at every boundary point of \(X_C\). Hence
\[
\begin{aligned}
    \int_{K_C}\widetilde f\,{\rm d}\overline{\nu}_C
    =
    \int_{K_I}
    \widetilde f\circ\overline{\pi}_C
    \,{\rm d}\overline{\nu}
    & =
    \lim_\alpha
    \int_{K_I}
    \widetilde f\circ\overline{\pi}_C
    \,{\rm d}\overline{\nu}_\alpha
    \\ & =
    \lim_\alpha
    \int_{X_C}
    f\,{\rm d}(\pi_C)_{\#}\nu_\alpha
    =
    \int_{X_C}f\,{\rm d}\mu_C.
\end{aligned}
\]
It follows from the Riesz representation theorem that the restriction of
\(\overline{\nu}_C\) to \(X_C\) is precisely \(\mu_C\):
\[
    \overline{\nu}_C|_{X_C}
    =
    \mu_C.
\]
Since both \(\overline{\nu}_C\) and \(\mu_C\) have total mass one,
\[
    \overline{\nu}_C(K_C\setminus X_C)=0,
\]
or equivalently,
\(
    \overline{\nu} (\overline{\pi}_C^{-1}(X_C)) = 1
\)    
for every \(C\in\mathfrak{C}\).
Because the contexts cover \(I\),
\[
    X_I
    =
    \bigcap_{C\in\mathfrak{C}}
    \overline{\pi}_C^{-1}(X_C).
\]
As \(\mathfrak{C}\) is finite, it follows that
\[
    \overline{\nu}(X_I)=1.
\]
The restriction
\(
    \nu:=\overline{\nu}|_{X_I}
\)
is therefore a Radon probability measure on \(X_I\), since the restriction of a Radon measure to an open subspace is Radon. Moreover, for every
\(C\in\mathfrak{C}\),
\[
    (\pi_C)_{\#}\nu
    =
    \mu_C.
\]
Thus \(\nu\) is a global coupling of the limiting family.
\end{proof}

\begin{theorem}[Kellerer-type marginal criterion]
\label{thm:kellerer_lch}
Let \(I\), \(\mathfrak{C}\), and \((X_i)_{i\in I}\) be as above, and let
\[
    \mu_C\in\mathsf{P}_{\mathrm{R}}(X_C),
    \qquad
    C\in\mathfrak{C}.
\]
The following assertions are equivalent:
\begin{enumerate}[\rm(1)]
    \item\label{it:kellerer_lch1} The family
    \(
        \{\mu_C\}_{C\in\mathfrak{C}}
    \)
    is probabilistically compatible.

    \item\label{it:kellerer_lch2} For every family of functions
    \(
        f_C\in C_{\rm b}(X_C;\mathbb{R})
    \),
    \(
        C\in\mathfrak{C},   
    \)
    one has
    \[
        \sum_{C\in\mathfrak{C}}
        \int_{X_C} f_C\,{\rm d}\mu_C
        \leq
        \sup_{x\in X_I}
        \sum_{C\in\mathfrak{C}}
        f_C(\pi_Cx).
    \]

 \item\label{it:kellerer_lch3} For every family of functions
    \(
        f_C\in C_{0}(X_C;\mathbb{R})
    \),
    \(
        C\in\mathfrak{C},   
    \)
    one has
    \[
        \sum_{C\in\mathfrak{C}}
        \int_{X_C} f_C\,{\rm d}\mu_C
        \leq
        \sup_{x\in X_I}
        \sum_{C\in\mathfrak{C}}
        f_C(\pi_Cx).
    \]
\end{enumerate}
\end{theorem}

\begin{proof}
Assume first that a global coupling
\(
    \mu_I\in\mathsf{P}_{\mathrm{R}}(X_I)
\)
exists. Then, for every family
\(f_C\in C_{\rm b}(X_C;\mathbb{R})\),
\[
\begin{aligned}
    \sum_{C\in\mathfrak{C}}
    \int_{X_C}f_C\,{\rm d}\mu_C
    &=
    \sum_{C\in\mathfrak{C}}
    \int_{X_I}f_C\circ\pi_C\,{\rm d}\mu_I
    \\
    &=
    \int_{X_I}
    \sum_{C\in\mathfrak{C}}
    f_C(\pi_Cx)\,{\rm d}\mu_I(x)
    \leq
    \sup_{x\in X_I}
    \sum_{C\in\mathfrak{C}}
    f_C(\pi_Cx).
\end{aligned}
\]
This proves
\(
    \ref{it:kellerer_lch1}\Longrightarrow \ref{it:kellerer_lch2}.
\)
The implication
\(
     \ref{it:kellerer_lch2}\Longrightarrow \ref{it:kellerer_lch3}
\)
is immediate from
\(
    C_0(X_C;\mathbb{R})
    \subseteq
    C_{\rm b}(X_C;\mathbb{R}).
\)

It remains to prove
\(
     \ref{it:kellerer_lch3}\Longrightarrow \ref{it:kellerer_lch1}.
\)
Let
\[
    \mathsf{E}
    :=
    \prod_{C\in\mathfrak{C}}\mathsf{M}(X_C)
\]
and define
\[
    T:
    \mathsf{P}_{\mathrm{R}}(X_I)
    \longrightarrow
    \mathsf{E},
    \qquad
    T\nu
    :=
    \bigl((\pi_C)_{\#}\nu\bigr)_{C\in\mathfrak{C}}.
\]
Put
\[
    \mathsf{K}
    :=
    T\bigl(\mathsf{P}_{\mathrm{R}}(X_I)\bigr).
\]
The set \(\mathsf K\) is convex, because \(\mathsf P_{\mathrm R}(X_I)\) is convex and \(T\) is affine. The space \(\mathsf E\), endowed with the product of the weak-* topologies, is a Hausdorff locally convex space.

Suppose, towards a contradiction, that
\(
    \mu
    :=
    (\mu_C)_{C\in\mathfrak{C}}
\)
does not admit a global coupling. Then
\(
    \mu\notin\mathsf{K}.
\)
Since every \(\mu_C\) is a Radon probability measure,
Lemma~\ref{lem:marginal_image_closed} implies that
\(
    \mu\notin\overline{\mathsf{K}}^{\,\mathsf E}.
\)
The Hahn--Banach separation theorem therefore gives a continuous real linear functional
\(
    \Lambda:\mathsf{E}\longrightarrow\mathbb{R}
\)
such that
\[
    \Lambda(\mu)
    >
    \sup_{\eta\in\overline{\mathsf{K}}^{\,\mathsf E}}
    \Lambda(\eta).
\]
Since \(\mathfrak{C}\) is finite and
\(
    \mathsf{M}(X_C)
    =
    C_0(X_C;\mathbb{R})^*
\)
with its weak-* topology, there exist functions
\(
    f_C\in C_0(X_C;\mathbb{R})
\), 
\(
    C\in\mathfrak{C}
\),
such that
\[
    \Lambda\bigl((\eta_C)_{C\in\mathfrak{C}}\bigr)
    =
    \sum_{C\in\mathfrak{C}}
    \int_{X_C}f_C\,{\rm d}\eta_C.
\]
Consequently,
\[
    \sum_{C\in\mathfrak{C}}
    \int_{X_C}f_C\,{\rm d}\mu_C
    >
    \sup_{\nu\in\mathsf{P}_{\mathrm{R}}(X_I)}
    \sum_{C\in\mathfrak{C}}
    \int_{X_C}f_C\,{\rm d}(\pi_C)_{\#}\nu.
\]
The supremum on the right equals
\[
\begin{aligned}
    \sup_{\nu\in\mathsf{P}_{\mathrm{R}}(X_I)}
    \int_{X_I}
    \sum_{C\in\mathfrak{C}}
    f_C(\pi_Cx)\,{\rm d}\nu(x)
    &=
    \sup_{x\in X_I}
    \sum_{C\in\mathfrak{C}}
    f_C(\pi_Cx),
\end{aligned}
\]
since every Dirac measure
\(\delta_x\), \(x\in X_I\), is a Radon probability measure. This contradicts condition~\emph{(3)}. Hence the family
\(\{\mu_C\}_{C\in\mathfrak{C}}\) admits a global coupling.
\end{proof}

\begin{remark}
When all outcome spaces \(X_i\) are finite, conditions~\ref{it:kellerer_lch2} and
\ref{it:kellerer_lch3} coincide and take the form
\[
    \sum_{C\in\mathfrak C}
    \sum_{x_C\in X_C}
    f_C(x_C)\mu_C(x_C)
    \leq
    \max_{x\in X_I}
    \sum_{C\in\mathfrak C}f_C(\pi_Cx).
\]
The resulting inequalities are precisely the separating inequalities
of the finite marginal polytope. In the Bell--CHSH application of
Section~\ref{sec:application_to_bell_chsh}, the relevant nontrivial members of this
family become the Fine--CHSH inequalities.  
\end{remark}

\begin{remark}
\label{rem:kellerer_generality}
Theorem~\ref{thm:kellerer_lch} is a locally compact Hausdorff version of the marginal duality criterion associated with Kellerer's work
\cite{Kellerer1964Marginal,Kellerer1984Duality}. Kellerer's results apply under substantially weaker topological and measurability assumptions. The present formulation is chosen because it admits a direct proof using compactification, weak-* compactness, and Hahn--Banach separation, while covering noncompact spaces such as \(\mathbb{R}^n\).
\end{remark}

\subsection{Operator-algebraic compatibility is strictly stronger than probabilistic compatibility}
\label{sec:gap_between_compatibility_conditions}

The relation between the two compatibility conditions is one-directional.
The positive implication is an immediate consequence of
Theorem~\ref{thm:global_classicalisation_commutative}.

\begin{corollary}
\label{cor:operator_implies_probabilistic}
If the family
\(
    \{\mathcal{M}_C\}_{C\in\mathfrak{C}}
\)
is operator-algebraically compatible, then, for every state
\(\rho\in\mathcal{S}(H)\), the associated family of contextwise laws
\(
    \{\mu_C^\rho\}_{C\in\mathfrak{C}}
\)
is probabilistically compatible.
\end{corollary}

\begin{proof}
By Proposition~\ref{prop:common-context-joint-pvm},
operator-algebraic compatibility is equivalent to commutativity of
\(\mathcal{M}_{\mathrm{env}}\).
Theorem~\ref{thm:global_classicalisation_commutative} then provides,
for every state \(\rho\), a global probability law
\(\mu_I^\rho\) satisfying
\[
    (\pi_C)_{\#}\mu_I^\rho
    =
    \mu_C^\rho
\]
for every \(C\in\mathfrak{C}\).
\end{proof}

The converse fails, already for two noncommuting qubit observables. In fact,
the following example shows the stronger fact that probabilistic compatibility
may hold separately for every state without the underlying PVMs being
operator-algebraically compatible.
The contexts in this first example are disjoint singleton contexts; this makes the probabilistic coupling automatic and isolates the distinction from overlap-consistency issues.

\begin{example}[Statewise coupling without operator compatibility]
\label{ex:probabilistic_without_operator_compatibility}
Let
\[
    H=\mathbb{C}^2,
    \qquad
    I=\{1,2\},
    \qquad
    \mathfrak{C}=\bigl\{\{1\},\{2\}\bigr\},
\]
and consider the Pauli observables
\[
    \sigma_z
    =
    \begin{pmatrix}
        1&0\\
        0&-1
    \end{pmatrix},
    \qquad
    \sigma_x
    =
    \begin{pmatrix}
        0&1\\
        1&0
    \end{pmatrix}.
\]
Let \(P_1\) and \(P_2\) be their spectral PVMs on
\(
    X_1=X_2=\{-1,1\},
\)
so that
\[
    P_1(\{p\})
    =
    \frac{1}{2}(I+p\sigma_z),
    \qquad
    P_2(\{q\})
    =
    \frac{1}{2}(I+q\sigma_x).
\]
The two prescribed context algebras
\[
    \mathcal{M}_{\{1\}}
    =
    W^*(\sigma_z),
    \qquad
    \mathcal{M}_{\{2\}}
    =
    W^*(\sigma_x)
\]
are individually commutative. Their enveloping algebra is, however,
\[
    \mathcal{M}_{\mathrm{env}}
    =
    W^*(\sigma_z,\sigma_x)
    =
    M_2(\mathbb{C}),
\]
which is noncommutative, since
\(
    [\sigma_z,\sigma_x]\neq 0.
\)
The two contexts are therefore not operator-algebraically compatible.

Nevertheless, let \(\rho\in\mathcal{S}(H)\) be arbitrary. The two
contextwise laws
\[
    \mu_{\{1\}}^\rho(p)
    =
    \operatorname{tr}\bigl(P_1(\{p\})\rho\bigr),
    \qquad
    \mu_{\{2\}}^\rho(q)
    =
    \operatorname{tr}\bigl(P_2(\{q\})\rho\bigr)
\]
always admit the product coupling
\[
    \mu_I^\rho(p,q)
    :=
    \mu_{\{1\}}^\rho(p)\,
    \mu_{\{2\}}^\rho(q),
    \qquad
    (p,q)\in\{-1,1\}^2.
\]
Indeed,
\[
    (\pi_{\{1\}})_{\#}\mu_I^\rho
    =
    \mu_{\{1\}}^\rho,
    \qquad
    (\pi_{\{2\}})_{\#}\mu_I^\rho
    =
    \mu_{\{2\}}^\rho.
\]
Thus the contextwise laws are probabilistically compatible for every
state \(\rho\), even though the original observable contexts are not
operator-algebraically compatible.
\end{example}

The strictness is not confined to disjoint contexts. In the Bell--CHSH cover of Section~\ref{sec:application_to_bell_chsh}, choose noncommuting sharp observables on each wing and take the maximally mixed state \(\rho=\tfrac14 I\). Every pair law is then the uniform law on \(\{-1,1\}^2\), so the four laws admit the uniform coupling on \(\{-1,1\}^4\), whereas operator-algebraic compatibility still fails.

The preceding example also clarifies why the existence of a probabilistic
coupling for every state does not recover operator-algebraic compatibility.
The couplings in the example are chosen separately for each state. In
particular,
\[
    \mu_I^\rho(p,q)
    =
    \operatorname{tr}\bigl(P_1(\{p\})\rho\bigr)
    \operatorname{tr}\bigl(P_2(\{q\})\rho\bigr),
\]
which is in general not affine in \(\rho\).

In fact, we have the stronger result that no affine choice of global couplings is possible in this
example. Suppose, to the contrary, that for every state \(\rho\) a coupling
\(\nu^\rho\) could be chosen such that, for every
\(p,q\in\{-1,1\}\), the mapping
\[
    \rho\longmapsto \nu^\rho(p,q)
\]
is affine. An affine real-valued function on the state space extends to an affine functional on the trace-one hyperplane of Hermitian matrices, and hence, by finite-dimensional trace duality, has the form \(\rho\mapsto\operatorname{tr}(G\rho)\) for a Hermitian matrix \(G\). If the function is nonnegative on all states, then \(G\geq0\), as is seen by testing on rank-one states. Thus there exist positive
operators \(G_{pq}\in M_2(\mathbb{C})\) such that
\[
    \nu^\rho(p,q)
    =
    \operatorname{tr}(G_{pq}\rho)
\]
for every state \(\rho\). Since \(\nu^\rho\) has the prescribed marginals
for every \(\rho\), it follows that
\[
    \sum_q G_{pq}
    =
    P_1(\{p\}),
    \qquad
    \sum_p G_{pq}
    =
    P_2(\{q\}).
\]
Hence
\[
    0\leq G_{pq}\leq P_1(\{p\}),
    \qquad
    0\leq G_{pq}\leq P_2(\{q\}).
\]
Since the marginal operators are projections, these inequalities imply that the range of \(G_{pq}\) is contained in each of the corresponding projection ranges; equivalently,
\[
    P_1(\{p\})G_{pq}
    =
    G_{pq}
    =
    G_{pq}P_1(\{p\}),
\]
and similarly for \(P_2(\{q\})\).

Using the two marginal identities and the orthogonality of the spectral
projections, we therefore obtain
\[
\begin{aligned}
    P_1(\{p\})P_2(\{q\})
    &=
    P_1(\{p\})\sum_{p'=\pm1}G_{p'q}
    =
    G_{pq},\\
    P_2(\{q\})P_1(\{p\})
    &=
    P_2(\{q\})\sum_{q'=\pm1}G_{pq'}
    =
    G_{pq}.
\end{aligned}
\]
Consequently,
\[
    P_1(\{p\})P_2(\{q\})
    =
    P_2(\{q\})P_1(\{p\})
\]
for all \(p,q\), contradicting the noncommutativity of the spectral PVMs
of \(\sigma_z\) and \(\sigma_x\).

Thus the distinction between the two compatibility notions is not merely that operator-algebraic compatibility is state-independent whereas probabilistic compatibility is formulated for a fixed state. What fails in the example is the possibility of choosing these couplings in a single affine, state-consistent way; for sharp measurements, such an affine choice would force the underlying PVMs to commute. Consequently, even the existence of a global probabilistic coupling separately for every state is strictly weaker than operator-algebraic compatibility. 

\subsection{Conclusion}

Global-context integration makes visible what is lost under classicalisation. 
At the operator-algebraic level, the prescribed measurement contexts admit a 
global integration only when the original observables are jointly compatible, 
equivalently, when their enveloping observable algebra is commutative. 
Classicalisation carries such an integration to a global coupling of the 
corresponding contextwise probability laws. In the passage from observables 
to probability laws, however, the commutation relations between observables 
belonging to different contexts are no longer retained. Consequently, the 
existence of a global probabilistic coupling cannot serve as an equivalent 
criterion for the existence of a common commutative observable context: 
classicalisation preserves global integration, but does not reflect it. 
The results of this section are summarised in 
Figure~\ref{fig:global_classicalisation_diagram}.

\begin{figure}[ht]
\centering
\resizebox{0.98\linewidth}{!}{%
\begin{tikzpicture}[
    node distance=2.8cm and 2.6cm,
    every node/.style={align=center},
    box/.style={draw, rectangle, minimum width=5.1cm, minimum height=1cm},
    >=latex
]

\node[box] (TL) {Prescribed commutative \\ subalgebras};
\node[box, right=of TL] (TR) {Contextwise probability \\ laws};
\node[box, below=of TL] (BL) {Enveloping observable \\ algebra};
\node[box, below=of TR] (BR) {Global-context probability \\ model};

\draw[->]
    (TL) -- node[above] {classicalisation} (TR);

\draw[->, dashed]
    (TL) -- node[left] {existence of a\\ global context} (BL);

\draw[->, dashed]
    (TR) -- node[right] {Kellerer criterion} (BR);

\draw[->]
    (BL) -- node[above] {classicalisation} (BR);

\end{tikzpicture}%
}
\caption{Relation between the operator-algebraic and probabilistic levels in the setting of global coupling.}
\label{fig:global_classicalisation_diagram}
\end{figure}

\section{Protocolwise integration}
\label{sec:protocol_wise_integration}

We now turn to the second interpretation of the integration problem. Rather
than identifying the prescribed contexts inside one common commutative
observable algebra, protocolwise integration retains the context label as
part of the integrated description. A protocol outcome records both the
context selected in a given run and the outcome obtained within that context.
At the operator-algebraic level, the different context algebras are
correspondingly retained as distinct summands of a direct-sum algebra. In particular, a measurement occurring in more than one prescribed context is represented by a separate copy in each corresponding summand; protocolwise integration does not impose a cross-context identification.

\subsection{Proof of the existence of a protocol integration}
Let
\(
    \lambda\in\mathsf{P}(\mathfrak{C})
\)
be a probability distribution with full support:
\[
    \lambda(C)>0
    \qquad
    \text{for every }C\in\mathfrak{C}.
\]
This distribution describes the random selection of the measurement context.

Define the protocol Hilbert space and protocol algebra by
\[
    H_{\mathrm{prot}}
    :=
    \bigoplus_{C\in\mathfrak{C}}H,
    \qquad
    \mathcal{M}_{\mathrm{prot}}
    :=
    \bigoplus_{C\in\mathfrak{C}}\mathcal{M}_C
    \subseteq
    \mathcal{L}(H_{\mathrm{prot}}).
\]
Since each \(\mathcal{M}_C\) is commutative and \(\mathfrak{C}\) is finite,
\(\mathcal{M}_{\mathrm{prot}}\) is a commutative von Neumann algebra.

The protocol outcome space is the disjoint union
\[
    X_{\mathrm{prot}}
    :=
    \bigsqcup_{C\in\mathfrak{C}}
    \bigl(\{C\}\times X_C\bigr),
\]
equipped with the \(\sigma\)-algebra
\[
    \Sigma_{\mathrm{prot}}
    :=
    \left\{
        \bigsqcup_{C\in\mathfrak{C}}
        \bigl(\{C\}\times E_C\bigr):
        E_C\in\bigotimes_{i\in C}\Sigma_i
        \text{ for every }C\in\mathfrak{C}
    \right\}.
\]
For \(E\in\Sigma_{\mathrm{prot}}\), write
\[
    E_C
    :=
    \{x\in X_C:(C,x)\in E\}.
\]

Fix a state \(\rho\in\mathcal{S}(H)\). Define the protocol state on
\(H_{\mathrm{prot}}\) by
\[
    \rho_{\mathrm{prot}}^{\lambda}
    :=
    \bigoplus_{C\in\mathfrak{C}}\lambda(C)\rho.
\]
This is a positive trace-class operator of trace \(\sum_C\lambda(C)\operatorname{tr}\rho=1\), and hence a state on \(H_{\mathrm{prot}}\).
Define
\[
    P_{\mathrm{prot}}:
    \Sigma_{\mathrm{prot}}
    \longrightarrow
    \operatorname{Proj}(H_{\mathrm{prot}})
\]
by
\[
    P_{\mathrm{prot}}(E)
    :=
    \bigoplus_{C\in\mathfrak{C}}P_C(E_C).
\]

We now prove the main result of this section. 
Although the construction is formally elementary, it is crucial to the central question explored here.

\begin{theorem}[Protocolwise integration]
\label{thm:protocol_wise_integration}
For every family of prescribed measurement contexts
\(
    \{\mathcal{M}_C\}_{C\in\mathfrak{C}},
\)
every state \(\rho\in\mathcal{S}(H)\), and every full-support context
distribution \(\lambda\), the following assertions hold:
\begin{enumerate}[\rm(1)]
    \item\label{it:protocol_wise_integration1} The mapping \(P_{\mathrm{prot}}\) is a PVM generating the
    commutative protocol algebra:
    \[
        \mathcal{M}_{\mathrm{prot}}
        =
        W^*\bigl(
            P_{\mathrm{prot}}(E):
            E\in\Sigma_{\mathrm{prot}}
        \bigr).
    \]

    \item\label{it:protocol_wise_integration2} Classicalisation of
    \((\mathcal{M}_{\mathrm{prot}},
    \rho_{\mathrm{prot}}^\lambda)\)
    with respect to \(P_{\mathrm{prot}}\) gives the probability law
    \[
        \mu_{\mathrm{prot}}^{\rho,\lambda}(E)
        :=
        \sum_{C\in\mathfrak{C}}
        \lambda(C)\mu_C^\rho(E_C),
        \qquad
        E\in\Sigma_{\mathrm{prot}}.
    \]

    \item\label{it:protocol_wise_integration3} Every prescribed contextwise law is recovered as a conditional
    law of the protocol model:
    \[
        \mu_C^\rho(E_C)
        =
        \frac{
            \mu_{\mathrm{prot}}^{\rho,\lambda}
            (\{C\}\times E_C)
        }{
            \mu_{\mathrm{prot}}^{\rho,\lambda}
            (\{C\}\times X_C)
        }
    \]
    for every \(C\in\mathfrak{C}\) and every
    \(E_C\in\bigotimes_{i\in C}\Sigma_i\).
\end{enumerate}
Consequently, protocolwise integration always exists.
\end{theorem}

\begin{proof}
The commutativity of \(\mathcal{M}_{\mathrm{prot}}\) follows directly
from the commutativity of its summands.

Since each \(P_C\) is a PVM, the direct-sum mapping
\[
    P_{\mathrm{prot}}(E)
    =
    \bigoplus_{C\in\mathfrak{C}}P_C(E_C)
\]
is projection-valued, normalised, and countably additive in the strong
operator topology. Hence \(P_{\mathrm{prot}}\) is a PVM.

Its range is contained in \(\mathcal{M}_{\mathrm{prot}}\). Conversely,
for every \(C\in\mathfrak{C}\) and every
\(E_C\in\bigotimes_{i\in C}\Sigma_i\),
\[
    P_{\mathrm{prot}}(\{C\}\times E_C)
    =
    (0,\ldots,0,P_C(E_C),0,\ldots,0).
\]
Since the range of \(P_C\) generates \(\mathcal{M}_C\), these projections,
as \(C\) varies, generate every summand of
\(\mathcal{M}_{\mathrm{prot}}\). Therefore
\[
    \mathcal{M}_{\mathrm{prot}}
    =
    W^*\bigl(
        P_{\mathrm{prot}}(E):
        E\in\Sigma_{\mathrm{prot}}
    \bigr).
\]

For
\[
    E
    =
    \bigsqcup_{C\in\mathfrak{C}}
    \bigl(\{C\}\times E_C\bigr),
\]
one has
\[
\begin{aligned}
    \operatorname{tr}\Bigl(
        P_{\mathrm{prot}}(E)
        \rho_{\mathrm{prot}}^\lambda
    \Bigr)
    &=
    \operatorname{tr}\left[
        \Bigl(
            \bigoplus_{C\in\mathfrak{C}}P_C(E_C)
        \Bigr)
        \Bigl(
            \bigoplus_{C\in\mathfrak{C}}\lambda(C)\rho
        \Bigr)
    \right]
    \\
    &=
    \sum_{C\in\mathfrak{C}}
    \lambda(C)\operatorname{tr}\bigl(P_C(E_C)\rho\bigr)
    =
    \sum_{C\in\mathfrak{C}}
    \lambda(C)\mu_C^\rho(E_C)
    =
    \mu_{\mathrm{prot}}^{\rho,\lambda}(E).
\end{aligned}
\]
Thus \(\mu_{\mathrm{prot}}^{\rho,\lambda}\) is precisely the probability
law obtained by classicalising the protocol algebra.

Finally,
\[
    \mu_{\mathrm{prot}}^{\rho,\lambda}
    (\{C\}\times X_C)
    =
    \lambda(C),
    \qquad 
    \mu_{\mathrm{prot}}^{\rho,\lambda}
    (\{C\}\times E_C)
    =
    \lambda(C)\mu_C^\rho(E_C).
\]
Since \(\lambda(C)>0\), taking the quotient gives
\[
    \mu_C^\rho(E_C)
    =
    \frac{
        \mu_{\mathrm{prot}}^{\rho,\lambda}
        (\{C\}\times E_C)
    }{
        \mu_{\mathrm{prot}}^{\rho,\lambda}
        (\{C\}\times X_C)
    }.
\]
\end{proof}

 We may summarise the protocolwise construction by the commuting diagram in
Figure~\ref{fig:protocol_classicalisation_diagram}.

\begin{figure}[ht]
\centering
\resizebox{0.98\linewidth}{!}{%
\begin{tikzpicture}[
    node distance=2.8cm and 2.6cm,
    every node/.style={align=center},
    box/.style={draw, rectangle, minimum width=5.1cm, minimum height=1cm},
    >=latex
]

\node[box] (TL) {Prescribed commutative \\ subalgebras};
\node[box, right=of TL] (TR) {Prescribed contextwise \\ laws};
\node[box, below=of TL] (BL) {Protocol \\ algebra};
\node[box, below=of TR] (BR) {Protocol level model};

\draw[->]
    (TL) -- node[above] {classicalisation} (TR);

\draw[->]
    (TL) -- node[left] {Direct-sum \\ integration} (BL);

\draw[->]
    (TR) -- node[right] {Protocol \\ mixture} (BR);

\draw[->]
    (BL) -- node[above] {classicalisation} (BR);

\end{tikzpicture}%
}
\caption{Relation between the operator-algebraic and probabilistic levels for protocol-level integration.}
\label{fig:protocol_classicalisation_diagram}
\end{figure}

\subsection{A common refinement of global and protocolwise integration}
\label{sec:common_refinement}

The two integration procedures introduced above differ in how they treat the
contextual structure. Protocolwise integration retains the context label,
whereas global integration removes the contextual separation by representing
all prescribed measurements in one common model. When global integration
exists at the operator-algebraic level, however, these two constructions are
not opposed: both can be obtained as reductions of a single larger
commutative model.

Assume throughout this subsection that the prescribed contexts are operator-algebraically compatible.
By
Proposition~\ref{prop:common-context-joint-pvm}, the enveloping algebra
\[
    \mathcal{M}_{\mathrm{env}}
    =
    W^*\Bigl(
        \bigcup_{C\in\mathfrak{C}}\mathcal{M}_C
    \Bigr)
\]
is commutative and the prescribed PVMs admit a joint PVM
\[
    P_I:
    \bigotimes_{i\in I}\Sigma_i
    \longrightarrow
    \operatorname{Proj}(H)
\]
on
\(
    X_I=\prod_{i\in I}X_i
\).
For a state \(\rho\), let
\[
    \mu_I^\rho(E)
    :=
    \operatorname{tr}\bigl(P_I(E)\rho\bigr)
\]
denote the corresponding global probability law.

We now adjoin the context variable to this global model. Define
\[
    H_{\mathrm{sup}}
    :=
    \bigoplus_{C\in\mathfrak{C}}H,
    \qquad
    \mathcal{M}_{\mathrm{sup}}
    :=
    \bigoplus_{C\in\mathfrak{C}}\mathcal{M}_{\mathrm{env}},
\]
and
\[
    \rho_{\mathrm{sup}}^\lambda
    :=
    \bigoplus_{C\in\mathfrak{C}}\lambda(C)\rho.
\]
The corresponding outcome space is
\[
    X_{\mathrm{sup}}
    :=
    \mathfrak{C}\times X_I
    =
    \bigsqcup_{C\in\mathfrak{C}}
    \bigl(\{C\}\times X_I\bigr).
\]
For a measurable set
\[
    E
    =
    \bigsqcup_{C\in\mathfrak{C}}
    \bigl(\{C\}\times E_C\bigr),
    \qquad
    E_C\in\bigotimes_{i\in I}\Sigma_i,
\]
define
\[
    P_{\mathrm{sup}}(E)
    :=
    \bigoplus_{C\in\mathfrak{C}}P_I(E_C).
\]
As in Theorem~\ref{thm:protocol_wise_integration},
\(P_{\mathrm{sup}}\) is a PVM, and its range generates
\(\mathcal{M}_{\mathrm{sup}}\). Classicalisation gives the probability law
\[
    \nu_{\mathrm{sup}}^{\rho,\lambda}(E)
    =
    \sum_{C\in\mathfrak{C}}
    \lambda(C)\mu_I^\rho(E_C).
\]
Equivalently,
\[
    \nu_{\mathrm{sup}}^{\rho,\lambda}
    =
    \lambda\otimes\mu_I^\rho.
\]

The global and protocolwise models can now both be recovered from this
common refinement.

\begin{proposition}[Common refinement of the two integrations]
\label{prop:common_refinement}
Under the assumptions above, define the measurable maps
\[
    G:X_{\mathrm{sup}}\longrightarrow X_I,
    \qquad
    G(C,x)=x,
\]
and
\[
    F:X_{\mathrm{sup}}\longrightarrow X_{\mathrm{prot}},
    \qquad
    F(C,x)
    =
    \bigl(C,\pi_Cx\bigr).
\]
Then
\[
    G_{\#}\nu_{\mathrm{sup}}^{\rho,\lambda}
    =
    \mu_I^\rho
\]
and
\[
    F_{\#}\nu_{\mathrm{sup}}^{\rho,\lambda}
    =
    \mu_{\mathrm{prot}}^{\rho,\lambda}.
\]
At the operator-algebraic level, the corresponding reductions of
\(\mathcal{M}_{\mathrm{sup}}\) are respectively the diagonal copy
\[
    \Delta(\mathcal{M}_{\mathrm{env}})
    :=
    \Bigl\{
        \bigoplus_{C\in\mathfrak{C}}a
        :
        a\in\mathcal{M}_{\mathrm{env}}
    \Bigr\}
    \cong
    \mathcal{M}_{\mathrm{env}}
\]
and the protocol algebra
\[
    \mathcal{M}_{\mathrm{prot}}
    =
    \bigoplus_{C\in\mathfrak{C}}\mathcal{M}_C.
\]
Thus global integration is obtained by forgetting the selected context,
whereas protocolwise integration is obtained by retaining the context and
forgetting the coordinates belonging to measurements outside that context.
\end{proposition}

\begin{proof}
Let \(E\in\bigotimes_{i\in I}\Sigma_i\). Since
\[
    G^{-1}(E)
    =
    \bigsqcup_{C\in\mathfrak{C}}
    \bigl(\{C\}\times E\bigr),
\]
we obtain
\[
\begin{aligned}
    (G_{\#}\nu_{\mathrm{sup}}^{\rho,\lambda})(E)
    =
    \nu_{\mathrm{sup}}^{\rho,\lambda}
    \bigl(G^{-1}(E)\bigr)
    =
    \sum_{C\in\mathfrak{C}}
    \lambda(C)\mu_I^\rho(E)
    =
    \mu_I^\rho(E).
\end{aligned}
\]
Hence
\[
    G_{\#}\nu_{\mathrm{sup}}^{\rho,\lambda}
    =
    \mu_I^\rho.
\]
At the level of PVMs,
\[
    P_{\mathrm{sup}}\bigl(G^{-1}(E)\bigr)
    =
    \bigoplus_{C\in\mathfrak{C}}P_I(E).
\]
Since the range of \(P_I\) generates
\(\mathcal{M}_{\mathrm{env}}\), these projections generate precisely
\(\Delta(\mathcal{M}_{\mathrm{env}})\). The restriction of the super-state
to this diagonal algebra agrees with the original global state, since
\[
\begin{aligned}
    \operatorname{tr}\left[
        \Bigl(\bigoplus_C a\Bigr)
        \rho_{\mathrm{sup}}^\lambda
    \right]
    &=
    \sum_C\lambda(C)\operatorname{tr}(a\rho)
    \\
    &=
    \operatorname{tr}(a\rho).
\end{aligned}
\]
Thus the diagonal reduction is canonically equivalent to the global
observable model
\((\mathcal{M}_{\mathrm{env}},\rho)\).

For the protocol reduction, let
\[
    E
    =
    \bigsqcup_{C\in\mathfrak{C}}
    \bigl(\{C\}\times E_C\bigr)
    \in\Sigma_{\mathrm{prot}}.
\]
Then
\[
    F^{-1}(E)
    =
    \bigsqcup_{C\in\mathfrak{C}}
    \Bigl(
        \{C\}\times\pi_C^{-1}(E_C)
    \Bigr).
\]
Consequently,
\[
\begin{aligned}
    (F_{\#}\nu_{\mathrm{sup}}^{\rho,\lambda})(E)
    =
    \sum_{C\in\mathfrak{C}}
    \lambda(C)
    \mu_I^\rho\bigl(\pi_C^{-1}(E_C)\bigr)
    =
    \sum_{C\in\mathfrak{C}}
    \lambda(C)\mu_C^\rho(E_C)
    =
    \mu_{\mathrm{prot}}^{\rho,\lambda}(E),
\end{aligned}
\]
where the second equality follows from
Theorem~\ref{thm:global_classicalisation_commutative}. Hence
\[
    F_{\#}\nu_{\mathrm{sup}}^{\rho,\lambda}
    =
    \mu_{\mathrm{prot}}^{\rho,\lambda}.
\]

Likewise,
\[
\begin{aligned}
    P_{\mathrm{sup}}\bigl(F^{-1}(E)\bigr)
    =
    \bigoplus_{C\in\mathfrak{C}}
    P_I\bigl(\pi_C^{-1}(E_C)\bigr)
    =
    \bigoplus_{C\in\mathfrak{C}}
    P_C(E_C)
    =
    P_{\mathrm{prot}}(E).
\end{aligned}
\]
The corresponding restricted observable algebra is therefore exactly
\(\mathcal{M}_{\mathrm{prot}}\).
\end{proof}

The proposition shows that, whenever operator-algebraic global integration
exists, the distinction between global and protocolwise integration may be
understood as a distinction between two coarse-grainings of a common
commutative model:
\[
\begin{array}{ccccc}
&&
\bigl(
    X_{\mathrm{sup}},
    \nu_{\mathrm{sup}}^{\rho,\lambda}
\bigr)
&&
\\[2mm]
&
\swarrow{\scriptstyle F}
&&
\searrow{\scriptstyle G}
&
\\[2mm]
\bigl(
    X_{\mathrm{prot}},
    \mu_{\mathrm{prot}}^{\rho,\lambda}
\bigr)
&&&&
\bigl(
    X_I,
    \mu_I^\rho
\bigr).
\end{array}
\]
The left-hand reduction retains which context was selected but discards the
outcomes assigned to measurements outside that context. The right-hand
reduction retains the complete assignment of outcomes but discards which
context was selected. What distinguishes the two integrations is therefore
not that they cannot coexist, but which information is retained in passing
from the common refinement to the integrated description.

\subsection{Conclusion for protocolwise integration}

The essential feature of protocolwise integration is that it does not remove
the contextual distinction. Its integrated description retains both the
selected context and the outcome obtained within that context. Thus a protocol
outcome has the form
\(
    (C,x_C)\in X_{\mathrm{prot}}
\),
rather than a context-independent assignment
\(
    x_I\in X_I=\prod_{i\in I}X_i
\).
Likewise, at the operator-algebraic level, the direct-sum construction
\(
    \mathcal{M}_{\mathrm{prot}}
    =
    \bigoplus_{C\in\mathfrak{C}}\mathcal{M}_C
\)
retains the different context algebras as distinct central summands, rather
than identifying their common observables inside one commutative envelope.

The preceding subsection shows that whenever operator-algebraic global
integration exists, global and protocolwise integration arise as two
reductions of a common commutative refinement. Protocolwise integration
retains the selected context while discarding the coordinates belonging to
measurements outside that context, whereas global integration retains the
full context-independent assignment while discarding the context label. The asymmetry between the two constructions becomes apparent when global
compatibility fails. Protocolwise integration does not require either of the
compatibility conditions of
Section~\ref{sec:global-context_integration}: by keeping the contexts
distinguished, the construction of
Theorem~\ref{thm:protocol_wise_integration} remains available even when one
or both of the corresponding global compatibility conditions fail. It is thus
precisely the additional requirement that the contextwise descriptions be
identified within one context-independent model that gives rise to the global
compatibility problem.

\section{Application to the Bell--CHSH experiment}\label{sec:application_to_bell_chsh}

We now apply the two integration procedures developed in the preceding
sections to the Bell--CHSH experiment. The purpose of this section is to
identify the operator-algebraic and probabilistic compatibility conditions
of Section~\ref{sec:global-context_integration} and the protocolwise
construction of Section~\ref{sec:protocol_wise_integration} in the standard
Bell--CHSH setting. In particular, we show that where global integration
fails, protocolwise integration does not. This sets up the discussion in
the next section. An idealised experimental protocol is recalled in the Appendix; the corresponding quantum-circuit implementation is given in Section~\ref{sec:bell_chsh_protocol}.

\subsection{The Bell--CHSH contexts}
\label{sec:bell_chsh_contexts}

Let
\[
    H=H_A\otimes H_B,
    \qquad
    X:=\{-1,1\}.
\]
Alice has two binary PVMs
\[
    P_i^A:2^X\longrightarrow\operatorname{Proj}(H_A),
    \qquad i\in\{1,2\},
\]
and Bob has two binary PVMs
\[
    P_j^B:2^X\longrightarrow\operatorname{Proj}(H_B),
    \qquad j\in\{1,2\}.
\]
For \(p,q\in X\), we abbreviate
\[
    P_i^A(p):=P_i^A(\{p\}),
    \qquad
    P_j^B(q):=P_j^B(\{q\}).
\]

On \(H\), the four local PVMs are represented by
\[
    P_{A_i}(p):=P_i^A(p)\otimes I,
    \qquad
    P_{B_j}(q):=I\otimes P_j^B(q).
\]
In the notation of Section~\ref{sec:integration_problem}, put
\[
    I=\{A_1,A_2,B_1,B_2\},
    \qquad
    \mathfrak{C}
    =
    \{C_{ij}:i,j\in\{1,2\}\},
\]
where
\(
    C_{ij}:=\{A_i,B_j\}.
\)
Since operators belonging to different tensor factors commute, every
\(C_{ij}\) is a measurement context. Its joint PVM is
\[
    P_{ij}(p,q)
    :=
    P_i^A(p)\otimes P_j^B(q),
    \qquad
    (p,q)\in X^2,
\]
and the corresponding commutative observable algebra is
\[
    \mathcal{M}_{ij}
    :=
    W^*\bigl(
        P_i^A(p)\otimes I,\,
        I\otimes P_j^B(q):
        p,q\in X
    \bigr).
\]

For a state \(\rho\in\mathcal{S}(H)\), contextwise classicalisation gives
the four laws
\[
    \mu_{ij}^\rho(p,q)
    :=
    \operatorname{tr}\left[
        \bigl(P_i^A(p)\otimes P_j^B(q)\bigr)\rho
    \right].
\]
Their one-site marginals are consistent:
\[
    \sum_{q\in X}\mu_{ij}^\rho(p,q)
    =
    \operatorname{tr}\left[
        \bigl(P_i^A(p)\otimes I\bigr)\rho
    \right],
\]
which is independent of \(j\), and
\[
    \sum_{p\in X}\mu_{ij}^\rho(p,q)
    =
    \operatorname{tr}\left[
        \bigl(I\otimes P_j^B(q)\bigr)\rho
    \right],
\]
which is independent of \(i\).

Let
\[
    A_i
    :=
    \sum_{p=\pm1}pP_i^A(p),
    \qquad
    B_j
    :=
    \sum_{q=\pm1}qP_j^B(q)
\]
be the associated binary observables. The correlation in the context
\(C_{ij}\) is
\[
    E_{ij}^\rho
    :=
    \sum_{p,q=\pm1}pq\,\mu_{ij}^\rho(p,q)
    =
    \operatorname{tr}\left[
        (A_i\otimes B_j)\rho
    \right].
\]

\subsection{Global integration}
\label{sec:bell_chsh_global}

The two global compatibility conditions of
Section~\ref{sec:global-context_integration} now take the following form.

\begin{proposition}
\label{prop:bell_chsh_global}
In the setting described above, the following assertions hold.
\begin{enumerate}
    \item The family
    \(
        \{\mathcal{M}_{ij}\}_{i,j=1}^2
    \)
    is operator-algebraically compatible if and only if
    \[
        [A_1,A_2]=0
        \qquad\text{and}\qquad
        [B_1,B_2]=0.
    \]

    \item For a fixed state \(\rho\), the family
    \(
        \{\mu_{ij}^\rho\}_{i,j=1}^2
    \)
    is probabilistically compatible if and only if the CHSH-type inequality
    \[
        \sum_{i,j=1}^2s_{ij}E_{ij}^\rho
        \leq 2
    \]
    holds for every choice of signs
    \(
        s_{ij}\in\{-1,1\}
    \)
with 
    \(
        \prod_{i,j=1}^2s_{ij}=-1
    \).
\end{enumerate}
\end{proposition}

\begin{proof}
The enveloping observable algebra is
\[
    \mathcal{M}_{\mathrm{env}}
    =
    W^*\bigl(
        A_1\otimes I,\,
        A_2\otimes I,\,
        I\otimes B_1,\,
        I\otimes B_2
    \bigr).
\]
All operators associated with different parties commute. For a binary sharp observable \(A_i\), its spectral projections are \(\frac12(I\pm A_i)\), so \(W^*(P_i^A)=W^*(A_i)\), and similarly for Bob.
 Hence
\(\mathcal{M}_{\mathrm{env}}\) is commutative if and only if the two
Alice observables commute with each other and the two Bob observables
commute with each other. The first assertion follows from
Proposition \ref{prop:common-context-joint-pvm}.

For the second assertion, probabilistic compatibility means that there
exists a probability measure
\(
    \nu\in\mathsf{P}(X^4)
\)
whose \((A_i,B_j)\)-marginal is \(\mu_{ij}^\rho\) for every \(i,j\).
By Fine's theorem, the overlap-consistent family
\(\{\mu_{ij}^\rho\}_{i,j=1}^2\) admits such a coupling if and only if
the Fine--CHSH inequalities hold
\cite{Fine1982PRL,Fine1982JMP}. Here positivity and normalisation are already built into the four probability laws, while consistency of the one-site marginals was established above; under these conditions the stated family of CHSH inequalities is also sufficient.

The necessity of these inequalities also follows directly from
Theorem \ref{thm:kellerer_lch}. Indeed, for
\[
    f_{ij}(p,q):=s_{ij}pq,
\]
the Kellerer-type marginal condition of Theorem \ref{thm:kellerer_lch} gives
\[
    \sum_{i,j=1}^2s_{ij}E_{ij}^\rho
    \leq
    \max_{a_1,a_2,b_1,b_2=\pm1}
    \sum_{i,j=1}^2s_{ij}a_ib_j
    =
    2.
\]
Indeed, if \(t_{ij}:=s_{ij}a_ib_j\), then \(\prod_{i,j}t_{ij}=-1\), so an odd number of the four signs \(t_{ij}\) is negative and their sum is either \(2\) or \(-2\). If a given assignment yields \(-2\), replacing both \(a_1\) and \(a_2\) by their negatives changes the sign of all four terms and yields \(2\); hence the maximum is \(2\).

\end{proof}

For the standard quantum instance, let
\(
    H_A=H_B=\mathbb{C}^2
\)
and define
\[
    M_\theta
    :=
    \cos(2\theta)\sigma_z+\sin(2\theta)\sigma_x,
    \qquad
    P_\theta(p)
    :=
    \frac12(I+pM_\theta).
\]
Take
\[
    \rho_{\mathrm{Bell}}
    :=
    \lvert\Phi^+\rangle\langle\Phi^+\rvert,
    \qquad {\rm where } \ 
    \lvert\Phi^+\rangle
    :=
    \frac{1}{\sqrt2}
    \bigl(\lvert00\rangle+\lvert11\rangle\bigr).
\]
The one-site reduced states of \(\rho_{\mathrm{Bell}}\) are both \(\tfrac12 I\), and
\[
\begin{aligned}
 \operatorname{tr}\bigl((\sigma_z\otimes\sigma_z)\rho_{\mathrm{Bell}}\bigr)
 &=\operatorname{tr}\bigl((\sigma_x\otimes\sigma_x)\rho_{\mathrm{Bell}}\bigr)=1,\\
 \operatorname{tr}\bigl((\sigma_z\otimes\sigma_x)\rho_{\mathrm{Bell}}\bigr)
 &=\operatorname{tr}\bigl((\sigma_x\otimes\sigma_z)\rho_{\mathrm{Bell}}\bigr)=0.
\end{aligned}
\]
It follows that, for settings \(a_i\) and \(b_j\),
\[
    \mu_{ij}^{\mathrm{Bell}}(p,q)
    =
    \frac14
    \left[
        1+pq\cos\bigl(2(a_i-b_j)\bigr)
    \right],
\]
and hence
\[
    E_{ij}^{\mathrm{Bell}}
    =
    \cos\bigl(2(a_i-b_j)\bigr).
\]

For
\[
    a_1=0,
    \qquad
    a_2=\frac{\pi}{4},
    \qquad
    b_1=\frac{\pi}{8},
    \qquad
    b_2=-\frac{\pi}{8},
\]
this gives
\[
    E_{11}^{\mathrm{Bell}}
    +
    E_{12}^{\mathrm{Bell}}
    +
    E_{21}^{\mathrm{Bell}}
    -
    E_{22}^{\mathrm{Bell}}
    =
    2\sqrt2.
\]
Therefore the four context laws are not probabilistically compatible.
Moreover,
\[
    [M_{a_1},M_{a_2}]\neq0,
    \qquad
    [M_{b_1},M_{b_2}]\neq0,
\]
so the four context algebras are not operator-algebraically compatible.

\subsection{Protocolwise integration}
\label{sec:bell_chsh_protocol}

\subsubsection{Existence of protocolwise integration}

We now apply the protocolwise construction of
Section~\ref{sec:protocol_wise_integration} to the Bell--CHSH experiment.
Let
\[
    \lambda=(\lambda_{ij})_{i,j=1}^2,
    \qquad
    \lambda_{ij}>0,
    \qquad
    \sum_{i,j=1}^2\lambda_{ij}=1,
\]
be the probability distribution according to which the setting pair is
chosen. The protocol outcome space is
\[
    X_{\mathrm{prot}}
    =
    \{1,2\}^2\times X^2.
\]
Thus a protocol outcome
\(
    (i,j,p,q)
\)
records both the selected setting pair \((i,j)\) and the two outcomes
\(p,q\in\{-1,1\}\) obtained in that branch of the experiment.

Theorem~\ref{thm:protocol_wise_integration} gives the protocol law
\[
    \mu_{\mathrm{prot}}^{\rho,\lambda}(i,j,p,q)
    :=
    \lambda_{ij}\mu_{ij}^\rho(p,q).
\]
The original contextwise laws are recovered by conditioning on the
selected setting pair:
\[
    \mu_{\mathrm{prot}}^{\rho,\lambda}
    \bigl(p,q\mid i,j\bigr)
    =
    \mu_{ij}^\rho(p,q).
\]
At the operator-algebraic level, the corresponding protocol algebra and
state are
\[
    \mathcal{M}_{\mathrm{prot}}
    =
    \bigoplus_{i,j=1}^2\mathcal{M}_{ij},
    \qquad
    \rho_{\mathrm{prot}}^\lambda
    =
    \bigoplus_{i,j=1}^2\lambda_{ij}\rho.
\]
Consequently, protocolwise integration exists for every state \(\rho\),
independently of whether the four Bell contexts admit a global integration.

\subsubsection{Quantum-circuit realisation of protocolwise integration}

For the standard Bell instance considered above, this abstract construction
has a direct quantum-circuit realisation, as shown in Figure~\ref{fig:chsh_quantum_circuit}. Let \(C_A\) and \(C_B\) be two
auxiliary qubits encoding Alice's and Bob's setting choices, respectively.
Writing their computational-basis outcomes as \(x,y\in\{0,1\}\), we identify
\[
    i=x+1,
    \qquad
    j=y+1.
\]
The two Hadamard gates in Figure~\ref{fig:chsh_quantum_circuit} prepare the
uniform setting distribution
\[
    \lambda_{ij}=\frac14.
\]
The Bell state \(\lvert\Phi^+\rangle\) is prepared on \(A\otimes B\), after
which the setting registers control the basis rotations implementing the
four measurement choices.

More explicitly, let
\[
    R_y(\varphi):=
    \exp\Bigl(-\frac{i\varphi}{2}\sigma_y\Bigr)
\]
and put
\[
    U_x
    :=
    R_y\Bigl(-\frac{x\pi}{2}\Bigr),
    \qquad
    V_y
    :=
    R_y\Bigl(-\frac{\pi}{4}+\frac{y\pi}{2}\Bigr).
\]
Since
\[
    R_y(-2\theta)^*\sigma_zR_y(-2\theta)
    =
    \cos(2\theta)\sigma_z+\sin(2\theta)\sigma_x
    =M_\theta,
\]
measurement in the computational basis following
\(R_y(-2\theta)\) implements measurement of \(M_\theta\).
The two values of \(U_x\) implement
\[
    a_1=0,
    \qquad
    a_2=\frac{\pi}{4},
\]
while the two values of \(V_y\) implement
\[
    b_1=\frac{\pi}{8},
    \qquad
    b_2=-\frac{\pi}{8}.
\]
In the circuit, the controlled \(R_y(-\pi/2)\) on Alice's qubit realises \(U_x\), while Bob's unconditional \(R_y(-\pi/4)\) followed by the controlled \(R_y(\pi/2)\) realises \(V_y\).

Immediately before the final measurements, and after canonically reordering the tensor factors as \(C_A\otimes C_B\otimes A\otimes B\), the circuit state can therefore
be written as
\[
    \lvert\Psi_{\mathrm{prot}}\rangle
    =
    \frac12
    \sum_{x,y=0}^1
    \lvert x,y\rangle_{C_A C_B}
    \otimes
    (U_x\otimes V_y)\lvert\Phi^+\rangle_{AB}.
\]
If \(r,s\in\{0,1\}\) denote the computational-basis outcomes on \(A\) and
\(B\), respectively, then
\[
\begin{aligned}
    \mathbb{P}(x,y,r,s)
    &=
    \frac14
    \left|
        \langle r,s|
        (U_x\otimes V_y)
        |\Phi^+\rangle
    \right|^2
    =
    \frac14\,
    \mu_{x+1,y+1}^{\mathrm{Bell}}
    \bigl((-1)^r,(-1)^s\bigr).
\end{aligned}
\]
Upon setting
\[
    p=(-1)^r,
    \qquad
    q=(-1)^s,
\]
this is precisely
\[
    \mu_{\mathrm{prot}}^{\mathrm{Bell},\lambda}(i,j,p,q)
    =
    \lambda_{ij}\mu_{ij}^{\mathrm{Bell}}(p,q),
    \qquad
    \lambda_{ij}=\frac14.
\]
The circuit therefore realises the protocol measure of
Theorem~\ref{thm:protocol_wise_integration} as the Born probability law of
one enlarged quantum measurement procedure.

\begin{figure}[ht]
\centering
\begin{quantikz}[row sep=0.35cm, column sep=0.45cm]
\lstick{$C_A$}
    & \ket{0}
    & \gate{H}
    & \qw
    & \ctrl{1}
    & \qw
    & \meter{} \rstick{$x$}
\\
\lstick{$A$}
    & \ket{0}
    & \gate{H}
    & \ctrl{1}
    & \gate{R_y(-\pi/2)}
    & \qw
    & \meter{} \rstick{$r$}
\\
\lstick{$B$}
    & \ket{0}
    & \qw
    & \targ{}
    & \gate{R_y(-\pi/4)}
    & \gate{R_y(\pi/2)}
    & \meter{} \rstick{$s$}
\\
\lstick{$C_B$}
    & \ket{0}
    & \gate{H}
    & \qw
    & \qw
    & \ctrl{-1}
    & \meter{} \rstick{$y$}
\end{quantikz}
\caption{
Quantum-circuit realisation of the protocolwise Bell--CHSH model.
The registers \(C_A\) and \(C_B\) encode the uniformly distributed setting
bits \(x,y\in\{0,1\}\), corresponding to
\(i=x+1\) and \(j=y+1\).
The Bell pair is prepared on \(A\otimes B\), and the setting registers
control the basis rotations implementing the four Bell measurement
contexts. The measurement outcomes \(r,s\in\{0,1\}\) correspond to the
\(\{\pm1\}\)-valued outcomes
\(p=(-1)^r\) and \(q=(-1)^s\).
The joint Born law of the four recorded bits is
\(\mu_{\mathrm{prot}}^{\mathrm{Bell},\lambda}\) with
\(\lambda_{ij}=1/4\).
}
\label{fig:chsh_quantum_circuit}
\end{figure}

In the Heisenberg picture, after transferring the branch-dependent rotations from the states to the measured observables, the circuit also makes the direct-sum structure of the protocol algebra explicit. The setting-register decomposition gives a canonical
identification
\[
    \mathbb{C}^2\otimes\mathbb{C}^2\otimes H
    \cong
    \bigoplus_{x,y=0}^1 H,
\]
under which the protocol algebra may be represented as the block-diagonal
algebra
\[
    \mathcal{M}_{\mathrm{prot}}
    \cong
    \left\{
        \sum_{x,y=0}^1
        |x,y\rangle\langle x,y|
        \otimes M_{xy}
        :
        M_{xy}\in\mathcal{M}_{x+1,y+1}
    \right\}.
\]
The setting projections
\[
    Z_{xy}
    :=
    |x,y\rangle\langle x,y|\otimes I
\]
are central projections relative to this block-diagonal protocol algebra and select the four branches of the protocol. Thus the auxiliary setting registers do not merely generate the
classical random choice appearing in the probability law; they provide a
concrete realisation of the four central summands in the protocol algebra.

\subsubsection{Coherent control and the direct-sum description}

There is, however, a subtle difference between the abstract direct-sum
model and its circuit realisation. In the direct-sum construction, the
setting pair is represented as a classical alternative: the state
\[
    \rho_{\mathrm{prot}}^\lambda
    =
    \bigoplus_{i,j=1}^2 \lambda_{ij}\rho
\]
contains no coherence between the four setting branches. In
Figure~\ref{fig:chsh_quantum_circuit}, by contrast, the setting registers
\(C_A\) and \(C_B\) are themselves quantum systems. After the Hadamard
gates, the four setting pairs occur in a coherent superposition, and the
state \(\lvert\Psi_{\mathrm{prot}}\rangle\) therefore contains
off-diagonal terms between different values of \((x,y)\).

These coherences are not visible, however, to the protocol algebra. Indeed, if
\[
    \Delta_{\mathrm{set}}(\omega)
    :=
    \sum_{x,y=0}^1
    Z_{xy}\omega Z_{xy}
\]
denotes dephasing in the setting basis, then every
\(M\in\mathcal{M}_{\mathrm{prot}}\) is block diagonal with respect to the
same decomposition and hence
\[
    \operatorname{tr}(M\omega)
    =
    \operatorname{tr}\bigl(
        M\Delta_{\mathrm{set}}(\omega)
    \bigr).
\]
Thus, as long as the setting registers are read only in the basis
\(\{|x,y\rangle\}\) and the different setting branches are not coherently
recombined, the coherent circuit and the corresponding incoherent mixture
produce exactly the same protocol statistics. The distinction could in
principle be detected by measurements containing off-diagonal terms
between different setting sectors, or equivalently by first recombining
the setting branches and observing interference. Such measurements lie
outside \(\mathcal{M}_{\mathrm{prot}}\). Relative to the prescribed
protocol algebra, the coherent and dephased circuit states are therefore
indiscernible in precisely the sense of the reduction discussed in
Section~\ref{sec:previous-results}.

After dephasing, the circuit state is block diagonal:
\[
    \Delta_{\mathrm{set}}
    \bigl(
        |\Psi_{\mathrm{prot}}\rangle
        \langle\Psi_{\mathrm{prot}}|
    \bigr)
    =
    \frac14
    \sum_{x,y=0}^1
    |x,y\rangle\langle x,y|
    \otimes
    (U_x\otimes V_y)
    \rho_{\mathrm{Bell}}
    (U_x\otimes V_y)^* .
\]
This is the Schr\"odinger-picture form of the corresponding direct-sum
description. For every observable \(N\) on the Bell-pair system and every
setting pair \((x,y)\),
\[
\begin{aligned}
    &\operatorname{tr}\left[
        N
        (U_x\otimes V_y)
        \rho_{\mathrm{Bell}}
        (U_x\otimes V_y)^*
    \right]
     =
    \operatorname{tr}\left[
        (U_x\otimes V_y)^*
        N
        (U_x\otimes V_y)
        \rho_{\mathrm{Bell}}
    \right].
\end{aligned}
\]
Thus the branch-dependent rotations may equivalently be transferred from
the state to the measurement operators. This gives the Heisenberg-picture
description used in Theorem~\ref{thm:protocol_wise_integration}, in which
the state is
\[
    \bigoplus_{i,j=1}^2
    \frac14\,\rho_{\mathrm{Bell}}
\]
and the four summands carry the corresponding context observables
\(\mathcal{M}_{ij}\).

It is useful to distinguish the two equivalences involved here. The passage
from the coherent circuit state to its dephased version is an
indiscernibility relation relative to the fixed protocol algebra
\(\mathcal{M}_{\mathrm{prot}}\): the state changes while the prescribed
observables are held fixed. By contrast, the passage from the dephased
Schr\"odinger-picture description to the direct-sum Heisenberg-picture
description transforms the state and the observables simultaneously. It is
therefore a blockwise unitary equivalence of state--observable descriptions,
rather than a further instance of indiscernibility relative to one fixed
algebra. The two descriptions nevertheless yield the same protocol
probability law, and their Holevo descriptions correspond under the induced
unitary identification.

\subsection{Conclusion}

For the standard Bell--CHSH instance, the two forms of integration
separate sharply. As shown in the preceding subsection,
\(\mathcal{M}_{\mathrm{env}}\) is noncommutative and \(\{\mu_{ij}^{\mathrm{Bell}}\}_{i,j=1}^2\) has no global coupling,
whereas
\(\mu_{\mathrm{prot}}^{\mathrm{Bell},\lambda}\)
exists and is realised by the circuit above. Thus the prescribed Bell contexts fail operator-algebraic global integration, and their four conditional laws fail probabilistic global integration, while the same laws remain the conditional branches of one well-defined protocol model.

\section{Discussion}\label{sec:discussion}

In his work \textit{Probabilistic and Statistical Aspects of Quantum Theory} \cite{holevo2011probabilistic}, Alexander Holevo developed an operational approach that may be understood as a form of statistical compression. This approach is motivated by the conviction that the operator formalism of quantum mechanics requires clarification not primarily through the logical structure of propositions, but through the statistical structure of measurement. In classical mechanics and probability theory, functions have a transparent role: they assign values, or random variables, on an underlying state space. In quantum mechanics, by contrast, the role of operators is less immediately intuitive, since their physical significance is mediated by the probabilities they assign to possible measurement outcomes. Holevo's aim in \cite{holevo2011probabilistic} is therefore to display the origin of the basic elements of the operator formalism from the statistical analysis of experimental situations: preparations determine states, measurements determine probability laws, and the mathematical structure of quantum theory is to be understood through the relation between the two.

The central formal device in this approach is Holevo's notion of reduction.
A statistical model describes states only through the probability laws
generated by a prescribed class of measurements. Once this class is
restricted, distinctions between states that cannot be detected by the
admissible measurements disappear from the model: states that were distinct
in a more comprehensive description become statistically indistinguishable.
Reduction may therefore be understood as a form of statistical compression,
in which only the information accessible through the chosen measurements is
retained. Holevo proves, under suitable finite-dimensional assumptions, that
every separated statistical model with finitely valued measurements can be
obtained as the reduction of a classical model equipped with a restricted
class of measurements
\cite[Theorem~1.7.1]{holevo2011probabilistic}; this result was extended to
PVMs on separable Hilbert spaces in
\cite[Theorem~5.4]{van2026indiscernibility}. Reduction as such is not peculiar
to quantum mechanics: classical measurements likewise disclose only part of
the information contained in a state. The decisive difference concerns the
way in which these reductions can be combined. In a classical Kolmogorov model in which the observables under consideration are jointly defined, the different measurements may be regarded as compatible reductions of one underlying sample space, and hence as partial descriptions subordinate to a single joint description.
Quantum reductions are not in general organised
in this way. Each compatible family of sharp measurements determines its own
classical statistical description, but noncommuting families need not be
recoverable as reductions of one common maximal measurement. The distinctive
feature of quantum theory is therefore not the occurrence of statistical
compression, but its context-bound structure: the available classical
descriptions are organised by different, potentially incompatible measurement
contexts rather than by a single underlying classical phase space.

In the final addendum on hidden-variables, Holevo brings this point into direct contact with Bell's theorem. Holevo argues that the usual opposition between deterministic classical mechanics and probabilistic quantum mechanics is too crude: the relevant question is not simply whether randomness can be explained by underlying variables, but what kind of classical statistical reconstruction is being required \cite{holevo2011probabilistic}. In the final part of the supplement, where Holevo turns to composite systems, the EPR paradox, and Bell's inequality, this question becomes especially delicate. For a fixed pair of local observables there is no obstruction to a classical probability representation, since observables acting on different tensor factors commute. The obstruction appears when one asks for a representation that is stable across several possible choices of local observables and that also satisfies the locality or separability requirement appropriate to the EPR setting. Holevo then recovers the standard Bell conclusion: for suitable entangled states, the quantum correlations violate the inequalities satisfied by such local hidden-variable models.

Our aim has been to refine this account in two respects. First, following
\cite{van2026indiscernibility}, we formulate the integration problem for
general families of quantum observables rather than for the Bell experiment
alone. This abstraction brings into view two notions of compatibility that
apply to different objects. In Holevo's statistical framework, compatibility
is a property of measurements: a family of measurements is compatible when
all its members are subordinate to one parent measurement. For sharp quantum
measurements, this is equivalent to commutativity (as in Proposition~\ref{prop:common-context-joint-pvm}) and hence corresponds, in
the present terminology, to operator-algebraic compatibility.

This notion is appropriate when one asks whether the original measurements
belong to one common observable structure, but it is stronger than what is
required for a joint representation of their probability laws. In the
Bell--CHSH setting, the relevant marginal problem concerns probabilistic
compatibility: whether, for a fixed state, the four contextwise laws can be
realised as marginals of one probability measure. Bell-inequality violations
show that this need not be possible for suitable states and measurement
settings. The relation between the two notions is one-directional.
Operator-algebraic compatibility implies probabilistic compatibility for
every state, but the converse fails. As
Example~\ref{ex:probabilistic_without_operator_compatibility} shows, the
contextwise laws may admit a joint coupling for every state even when the underlying
observable contexts cannot be embedded into one common commutative algebra.
Compatibility must therefore be specified relative to the objects under
consideration: measurements may be compatible as parts of one observable
structure, while their reduced probability laws may be compatible as
marginals of one measure.

Second, the distinction between global and protocolwise integration shows
that the integration problem presupposes a prior individuation of the
experiment. Operationally, the Bell--CHSH experiment may be described as one
enlarged measurement procedure whose complete outcome records both the
selected settings and the two results obtained. On this description, repeated
runs realise one and the same protocol context, while the four
setting-conditioned laws arise only by conditioning on the values of the
setting variables. Protocolwise integration formalises precisely this description: the setting choice is incorporated into the outcome space of the experiment itself, and the different setting pairs appear as conditional branches of one larger context.

To describe the same procedure instead as a family of four measurement
contexts introduces an additional decomposition of this protocol-level
description. The four conditional branches are now treated as distinct
contexts, while occurrences of the same local measurement in different
branches are identified with one another. Thus Alice's measurement \(A_i\) in
\(C_{i1}\) is identified with \(A_i\) in \(C_{i2}\), and similarly on Bob's
side. This is precisely the structure presupposed by the mathematical
formulation of the integration problem in
Section~\ref{sec:integration_problem}, where a common index set \(I\) and a
family \(\mathfrak{C}\subseteq 2^I\) are fixed from the outset. The
mathematical question of global integration therefore arises only after this
decomposition of the protocol into overlapping contexts has been specified.

This leads to the central philosophical point: the distinction between global
and protocolwise integration brings into view the structure imposed on a
classical representation. Theorem~\ref{thm:protocol_wise_integration} shows
that, under its stated conditions, the protocol-level representation is always
available: a finite setting-dependent experiment may be represented as one
enlarged protocol context by including the setting variables among its recorded
outcomes. A Bell-local hidden-variable representation, by contrast, imposes
the stronger requirement that the contextwise descriptions admit a single
context-independent representation preserving the relevant cross-context
identifications \cite{brunner2014bell}. In the standard binary Bell--CHSH setting,  once measurement independence and the cross-context identity of each local response variable are imposed, Fine's theorem identifies this requirement with the existence of a global
coupling of the four contextwise laws. The physical significance of the failure
of global integration therefore lies not in an impossibility of classically
representing the experimental protocol as such, but in the incompatibility of the
observed correlations with this additional Bell-local structure.

In this sense, the distinction developed here returns to Holevo's question of
what kind of classical statistical reconstruction is being required. The
obstruction revealed by Bell--CHSH excludes a particular reconstruction in
which the contextwise descriptions are required to coexist in a single global
model preserving the prescribed cross-context structure. Seen from this
perspective, the relevant question is not simply whether a classical
statistical representation exists, but what structure that representation is
required to preserve. Global and protocolwise integration answer this question
differently. Their separation therefore makes explicit a structural choice
that precedes, and partly determines, the meaning of ``classical
representation'' itself.

\bigskip
{\em Acknowledgment and AI disclosure} -- The authors used versions 5.5 and 5.6 of the OpenAI ChatGPT-models as tools for mathematical brainstorming, restructuring, and editorial assistance while preparing the manuscript.  All arguments and the final text were checked and revised by the authors. The authors take full responsibility for the correctness of the results in the manuscript.

\bibliographystyle{plainurl}
\bibliography{Holevo_Classicalisation}

\appendix
\section{An idealised Bell--CHSH protocol}
\label{sec:experimental_setup}

Here we recall a Bell--CHSH protocol, which is an idealised version of the famous two-channel polariser experiment of Aspect, Grangier and Roger \cite{aspect1982experimental}. The independent random choice of a setting on every run in the description below is part of the present mathematical protocol and is not intended as a literal description of that static-polariser experiment.

\begin{figure}
    \centering
    \IfFileExists{Experimtental_setup_bell_overview.png}{%
        \includegraphics[width=0.8\linewidth]{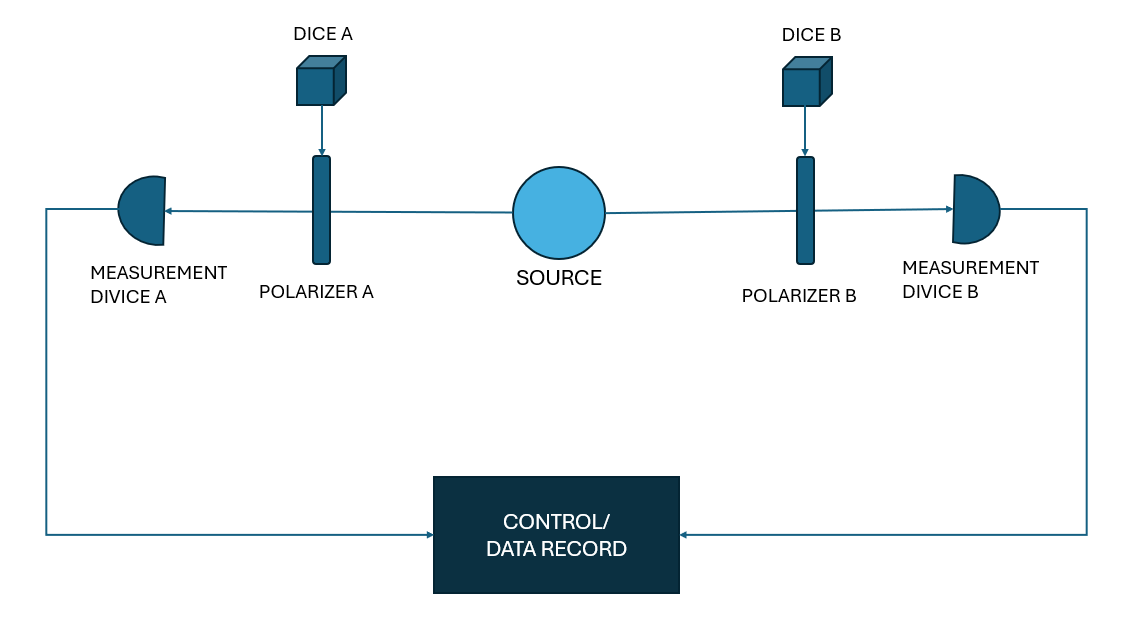}%
    }{%
        \fbox{\parbox[c][4cm][c]{0.76\linewidth}{\centering The external figure file was not supplied with the manuscript.}}%
    }
    \caption{A schematic overview of the Bell--CHSH experimental arrangement.}
    \label{fig:bell_set_up}
\end{figure}

A single run of the experiment proceeds as follows; see Figure
\ref{fig:bell_set_up}.

\begin{enumerate}
    \item[\rm(S1)] In the idealised two-qubit model, a source prepares a pair of photons in the Bell state
    \[
        \ket{h_{\rm Bell}}
        =\frac1{\sqrt2}\bigl(\ket0\ket0+\ket1\ket1\bigr),
    \]
    where \(\ket0,\ket1\) denote a fixed polarisation basis for each photon.
    The two photons propagate in opposite directions, towards Alice and Bob.

    \item[\rm(S2)] Alice's polariser setting is chosen from
    \(\{a_1,a_2\}\), and Bob's from \(\{b_1,b_2\}\).  The two choices are
    made independently, and each of the two local settings is selected with
    probability \(1/2\).  Thus each pair \((a_i,b_j)\) occurs with probability
    \(1/4\).\footnote{The present paper treats the setting choices as ordinary
    random variables, independent of the prepared photon pair and of each other.
    Nothing in the argument depends on whether these choices are implemented by
    classical randomisation or by a physical random-number generator.  The point
    of the assumption is that the selected settings are not already encoded in
    the state emitted by the source.}

    \item[\rm(S3)] The photons pass through the respective polarisers.

    \item[\rm(S4)] Alice and Bob record outcomes
    \(p,q\in\{-1,1\}\).  In the quantum description below these are represented
    by the spectral projections of rotated Pauli observables.
\end{enumerate}

After \(k\) runs of the experiment the data are grouped according to the chosen
settings.  For \((a,b)\in\{a_1,a_2\}\times\{b_1,b_2\}\), let
\(N_k^{(a,b)}\) denote the number of runs in which the settings were \((a,b)\).
Thus
\[
        k=N_k^{(a_1,b_1)}+N_k^{(a_1,b_2)}
          +N_k^{(a_2,b_1)}+N_k^{(a_2,b_2)}.
\]
For \((p,q)\in\{-1,1\}^2\), let \(N_k^{(a,b)}(p,q)\) denote the number of
runs in which the settings were \((a,b)\) and the recorded outcome was \((p,q)\).
Then
\[
        N_k^{(a,b)}
        =\sum_{p,q=\pm1}N_k^{(a,b)}(p,q).
\]
For those \(k\) for which \(N_k^{(a,b)}>0\), the empirical conditional
correlation at the setting pair \((a,b)\) is
\[
        E_k^{(a,b)}
        :=\frac1{N_k^{(a,b)}}
        \sum_{p,q=\pm1}pq\,N_k^{(a,b)}(p,q).
\]
The empirical CHSH expression is then
\[
        S_k(a_1,a_2,b_1,b_2)
        :=E_k^{(a_1,b_1)}+E_k^{(a_1,b_2)}
          +E_k^{(a_2,b_1)}-E_k^{(a_2,b_2)}.
\]
Assuming that successive runs are independent and identically distributed, and writing \(\mu_{ij}\) for the conditional outcome law at the setting pair \((a_i,b_j)\), the strong law of large numbers gives \(N_k^{(a_i,b_j)}/k\to1/4\) and, almost surely,
\[
        E_k^{(a_i,b_j)}
        \longrightarrow
        E^{(a_i,b_j)}
        :=\sum_{p,q=\pm1}pq\,\mu_{ij}(p,q).
\]
Accordingly, the limiting CHSH value is
\[
        S(a_1,a_2,b_1,b_2)
        :=E^{(a_1,b_1)}+E^{(a_1,b_2)}
          +E^{(a_2,b_1)}-E^{(a_2,b_2)}.
\]
The main text formulates this quantity directly in terms of the four fixed-setting probability laws.

With the standard CHSH convention used in this paper, the corresponding
ideal settings are
\[
        a_1=0,\qquad a_2=\frac\pi4,
        \qquad b_1=\frac\pi8,
        \qquad b_2=-\frac\pi8.
\]

Up to the corresponding relabelling of settings and outcomes, this is the CHSH convention used by Aspect et al. Their reported experimental value was
\[
        S_{\rm exp}=2.697\pm0.015
\]
\cite[pp.~93--94]{aspect1982experimental}.

\end{document}